\documentclass[5p, lefttitle, times]{elsarticle}
\usepackage[dvipsnames]{xcolor}
\usepackage{soul}
\usepackage{hyperref}
\usepackage{lineno}
\usepackage{booktabs}
\usepackage{multirow}
\usepackage{algorithmic,algorithm}
\usepackage{graphicx}
\usepackage{array}
\usepackage{todonotes}
\usepackage{gensymb}
\usepackage{amsmath}
\usepackage{amsfonts}
\usepackage{amssymb}
\usepackage{comment}
\usepackage{pgfplots}
\usepackage{appendix}

\usepackage{textcomp}
\usepackage{gensymb}

\usepackage[caption=false,font=footnotesize]{subfig}

\journal{Measurement}

\pgfplotsset{compat=1.16} 
 
\begin{document}
\begin{frontmatter}

\title{A novel method for error analysis in radiation thermometry with application to \\ industrial furnaces}
\cortext[correspondingauthor]{Corresponding author}

\author[VICOMTECH]{Iñigo Martinez}\corref{correspondingauthor}
\ead{imartinez@vicomtech.org}

\author[VICOMTECH]{Urtzi Otamendi}
\author[VICOMTECH]{Igor G. Olaizola}
\author[VICOMTECH]{Roger Solsona}
\author[VICOMTECH]{Mikel Maiza}
\author[TECNUN,ICDIA]{Elisabeth Viles}
\author[PETRONORINNOVACION]{Arturo Fernandez}
\author[PETRONOR]{Ignacio Arzua}

\address[VICOMTECH]{Vicomtech Foundation, Basque Research and Technology Alliance (BRTA), Donostia-San Sebastián 20009, Spain}

\address[TECNUN]{TECNUN School of Engineering, University of Navarra, Donostia-San Sebastián 20018, Spain}

\address[ICDIA]{Institute of Data Science and Artificial Intelligence, University of Navarra, Pamplona 31009, Spain}

\address[PETRONORINNOVACION]{Petronor Innovation, Muskiz 48550, Spain}
\address[PETRONOR]{Petróleos del Norte, Muskiz 48550, Spain}

\begin{abstract}
Accurate temperature measurements are essential for the proper monitoring and control of industrial furnaces. However, measurement uncertainty is a risk for such a critical parameter. 
Certain instrumental and environmental errors must be considered when using spectral-band radiation thermometry techniques, such as the uncertainty in the emissivity of the target surface, reflected radiation from surrounding objects, or atmospheric absorption and emission, to name a few.
Undesired contributions to measured radiation can be isolated using measurement models, also known as error-correction models.
This paper presents a methodology for budgeting significant sources of error and uncertainty during temperature measurements in a petrochemical furnace scenario.
A continuous monitoring system is also presented, aided by a deep-learning-based measurement correction model, to allow domain experts to analyze the furnace's operation in real-time.
To validate the proposed system's functionality, a real-world application case in a petrochemical plant is presented. The proposed solution demonstrates the viability of precise industrial furnace monitoring, thereby increasing operational security and improving the efficiency of such energy-intensive systems.
\end{abstract}

\begin{keyword}
radiation thermometry,
error analysis,
infrared imagery,
monitoring system,
petrochemical industry,
surrogate model,
deep learning
\end{keyword}

\end{frontmatter}

\section{Introduction}\label{sec:introduction}

\begin{figure*}[!htb]
  \centering
  \subfloat[]{\includegraphics[width=0.35\linewidth]{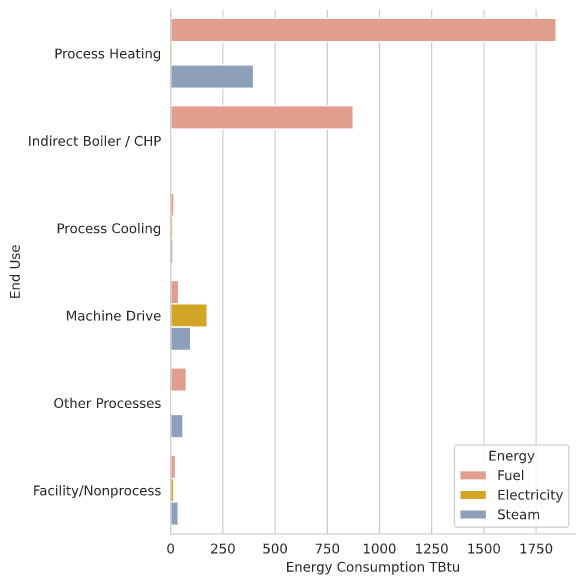}
    \label{fig:industryA}} 
  \subfloat[]{\includegraphics[width=0.63\linewidth]{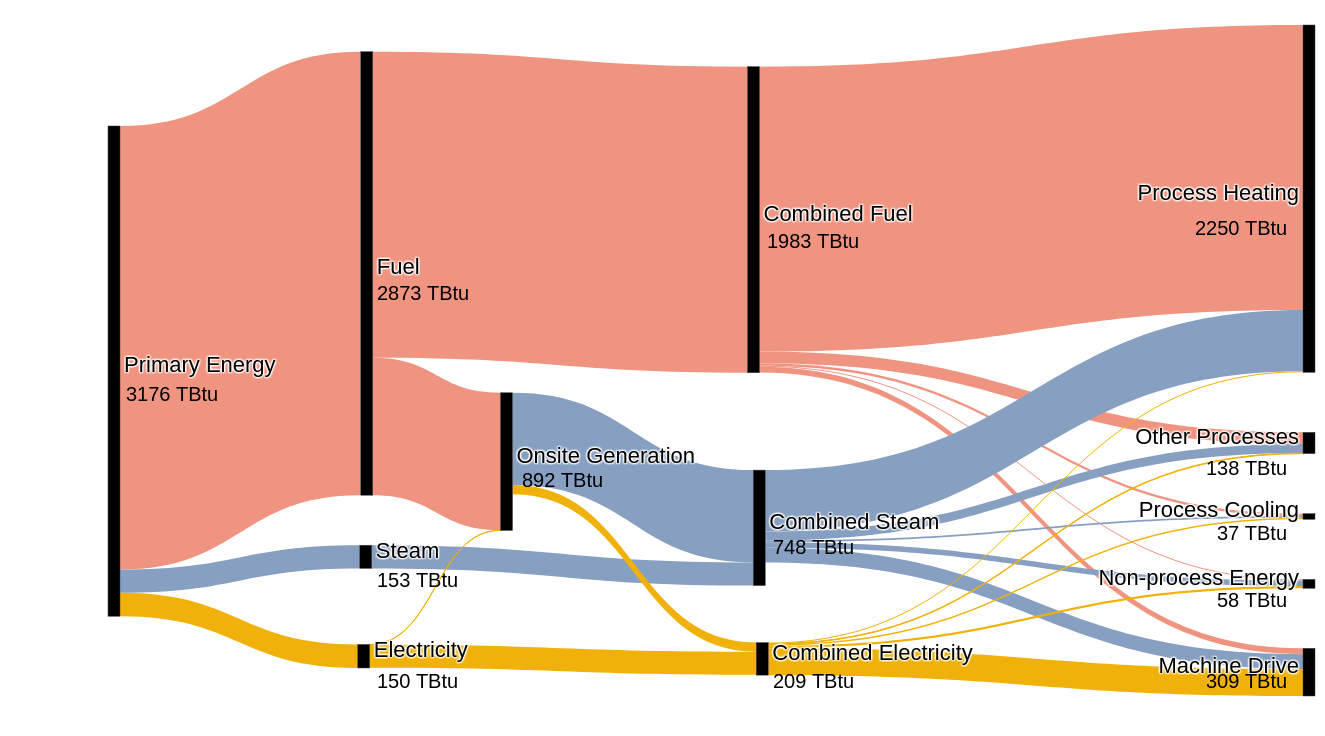}
    \label{fig:industryB}}
  \caption{Petrochemical Industry in the USA. a) Energy consumption by end-use, in TBtu (Trillion British thermal unit, 1 BTU = 1055 J).
   b) Sankey diagram: line widths indicate the volume of energy that flows to major energy end uses in petrochemical manufacturing and line colors indicate fuel, steam, electricity applied energy. Figures created from data by the U.S. Department of Energy (DOE) 2015 \cite{osti_1248754}.
  } 
  \label{fig:industry}
\end{figure*}

In crude oil refineries, chemical and petrochemical plants, heating processes account for up to 85\% of the overall energy consumption, as reported by the U.S. Department of Energy \cite{osti_1248754}.
These processes require almost 95\% of the fuel, 65\% of the steam, and 4\% of the electricity required for the processes of separation and chemical conversion of products. Fig. \ref{fig:industryA} summarizes the energy consumption by end-use in the U.S. petrochemical industry.

The furnace is a heat exchanger in which the process fluid flows through tubes and is heated by \textit{radiation} from a combustion flame that is generated by oxidizing fuel and by \textit{convection} from hot combustion gases. Process furnaces provide a specific amount of heat to the fluid being heated, at high-temperature levels, without causing localized overheating to the fluid or structural components. 

\begin{figure}[!htb]
    \centering
    \includegraphics[width=0.6\linewidth]{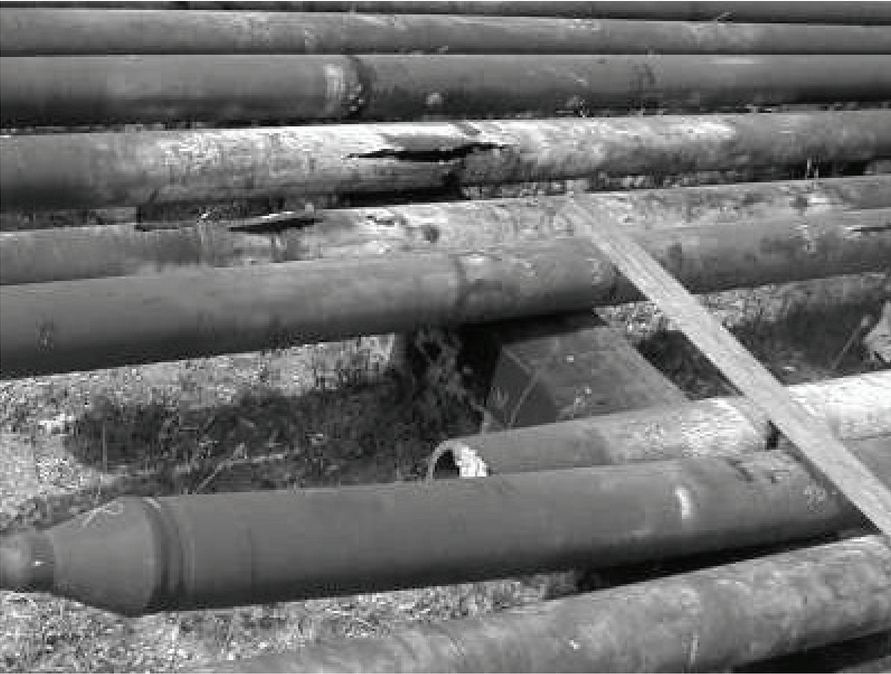} \\
    \vspace{1em}
    \includegraphics[width=0.6\linewidth]{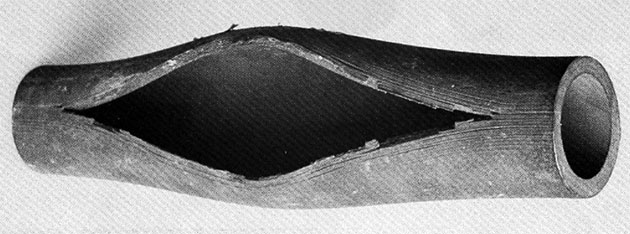}
    \caption{Damaged and ruptured tubes due to over-heating \cite{saunders, overheat}.}
    \label{fig:tubes}
\end{figure}

The energy efficiency of a furnace is critical in terms of sustainability. Because of the high temperatures and fuel liquids and gases involved, such processes must be continuously monitored to ensure safety and avoid material and human damages.
Accurate temperature measurements are thus relevant to control and monitor the internal temperature of industrial furnaces. Measurement uncertainty is a risk for such a critical parameter.
Through appropriate methods, accurate temperature data acquisition helps in characterizing the temperature distribution inside the furnace. 
Reliable measurements allow to: a) predict the \textit{Remaining Useful Life} (RUL) of furnace tubes, b) avoid catastrophic failures by managing hot tubes before they reach a critical state \cite{life_intro} (see Fig. \ref{fig:tubes}), c) optimize online production, extending run times, d) prevent unplanned outages and lost production time and e) make informed decisions about maintenance timing and coordination.

However, acquiring reliable and accurate temperature measurements inside an industrial furnace is a challenging task. The extreme atmospheric conditions of these furnaces limit the applicable measurement techniques and make it difficult to obtain accurate data. See section \ref{sec:thermometry} for a summary of currently available temperature measurement techniques.

Spectral-band radiation thermometry techniques are most widely used in such environments \cite{ir_intro, ir_intro2}. However, there are certain measurement and environmental errors that need to be taken into account: the uncertainty in the emissivity of the target surface, the reflected radiation from surrounding objects, or the atmospheric absorption and emission, among others \cite{saunders} (see Fig. \ref{fig:furnace_diagram} for a visualization of the energy exchange by radiation on a furnace tube). Undesired contributions to the measured radiation are isolated using measurement models, also called error-correction models.

Error-correction models are generally based on Planck's Law, which describes the spectral-energy distribution of radiation $G$ emitted by a blackbody at a given temperature $T$. Radiation thermometry techniques require applying the inverse of Planck's Law to infer the temperature of the target surface $T$ from a given radiation measurement $G$. Embedding the inverse of Planck's Law on error-correction models leads to an unconstrained optimization problem, which may be computationally expensive to solve. This procedure may limit the potential applications of accurate spectral-band radiation thermometry.
To address this limitation, surrogate models, high-fidelity models, are a potential solution.

This article introduces a methodology to identify and quantify primary sources of error of spectral-band radiation thermometry in a petrochemical furnace scenario. 
The result of such analysis yields an error-correction model that includes the most critical errors. To integrate this model into a continuous and real-time monitoring system, we propose a scalable data architecture solution, and we developed a surrogate model to speed up the measurement error-correction procedure.

Our main contributions are the following: first, we present a study of the error and uncertainty in spectral-band radiation thermometry for petrochemical furnaces. Then, we introduce four different mathematical models to correct temperature measurements in increasing order of complexity. Second, we integrate these error-correction models into a continuous monitoring system through a deep-learning-based surrogate model. In addition, we propose a computing architecture with user-friendly interaction interfaces to enable domain experts to analyze the furnace's operation in real-time. Finally, we present a real application case in a petrochemical plant to validate the proposed system's functionality.

\newpage

The remainder of the article is structured as follows:
Section \S\ref{sec:thermometry} provides an overview of temperature measurement methods.
Section \S\ref{sec:uncertainty_budget} presents the error analysis of spectral-band radiation thermometry in a petrochemical furnace.
Section \S\ref{sec:monitoring_system} introduces the thermal imagery acquisition methodology and the architectural design of the monitoring system.
Finally, section \S\ref{sec:conclusions} contains the concluding remarks and outlines some areas for further research.

\usetikzlibrary{shapes}
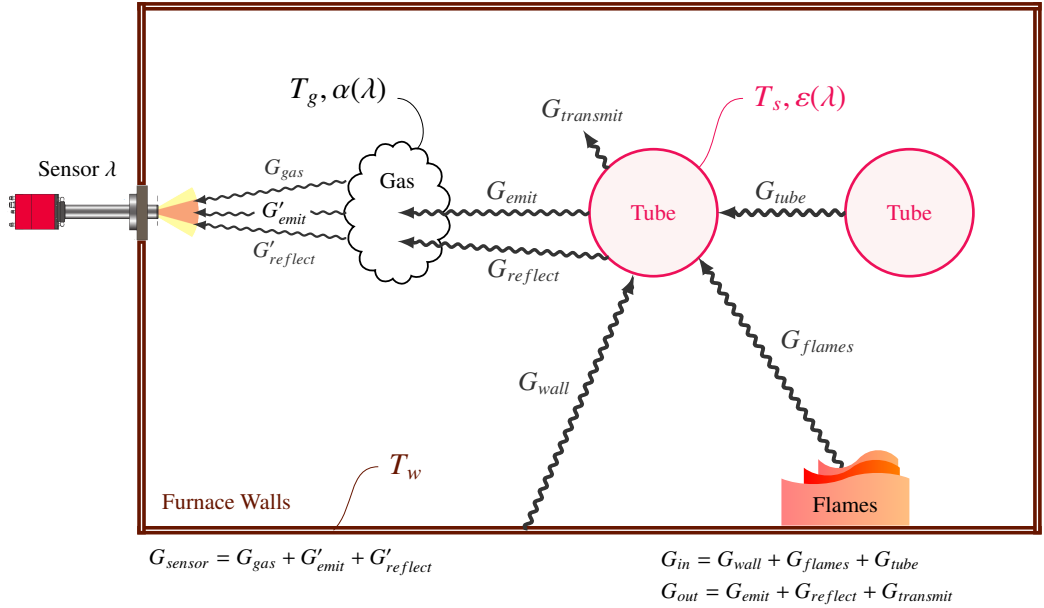
\begin{figure*}[!htb]
\hspace{1.5cm}
\resizebox{0.75\linewidth}{!}{ 
    \begin{tikzpicture}[shift={(0,0)}]
    \pgfmathsetmacro{\n}{14}
    \pgfmathsetmacro{\m}{8.25}
    \pgfmathsetmacro{\p}{5}
    \pgfmathsetmacro{\h}{0.5}
    \pgfmathsetmacro{\s}{1.35}


    \draw[Sepia, ultra thick, double, line cap=rect] (0,0) -- (\n, 0);
    \draw[Sepia, ultra thick, double, line cap=rect] (\n, 0) -- (\n, \m);
    \draw[Sepia, ultra thick, double, line cap=rect] (\n, \m) -- (0, \m);
    \draw[Sepia, ultra thick, double, line cap=rect] (0,0) -- (0, \p - \h);
    \draw[Sepia, ultra thick, double, line cap=rect] (0,\m) -- (0, \p + \h);
    
    \begin{scope}[shift={(3, 0)}]
        \draw[Sepia] (0, 0) to[out=45,in=180] (0.7,1) node[anchor=west, scale=\s] {$T_w$};
    \end{scope}
    \node[Sepia, anchor=south west] at (0.2,0.2) {Furnace Walls};

    \begin{scope}[shift={(4, \p)}]
        \node [cloud, draw, cloud puffs=15, cloud puff arc=140, 
        aspect=1.5, rotate=90, inner ysep=0.7em, scale=2, fill=white, thick] at (0,0) {};
        \draw[black] (0.3, 1) to[out=45,in=0] (0,1.8) node[anchor=base east, scale=\s] {$T_g, \alpha(\lambda)$};
        \node[black] at (0, 0.5) {Gas};
    \end{scope}

    \begin{scope}[shift={(8, \p)}]
        \filldraw[color=OrangeRed, fill=OrangeRed!5, very thick] (0,0) circle (1);
        \draw[OrangeRed] (0.7, 0.7) to[out=45,in=180] (1.4,1.8) node[anchor=west, scale=\s] {$T_s, \varepsilon(\lambda)$};
        \draw[-latex, solid, line width=1.7pt, black!80, decorate, 
        decoration={snake,amplitude=.4mm,segment length=2mm,post length=1mm}]
        (-0.7,0.7) -- node[pos=1.0, above] {\large $G_{transmit}$} (-1.1,1.3);
        \node[OrangeRed] {Tube};
    \end{scope}
    
    \begin{scope}[shift={(12, \p)}]
        \filldraw[color=OrangeRed, fill=OrangeRed!5, very thick] (0,0) circle (1);
        \draw[-latex, solid, line width=1.7pt, black!80, decorate, 
        decoration={snake,amplitude=.4mm,segment length=2mm,post length=1mm}]
        (-1,0) -- node[midway, above] {\large $G_{tube}$} (-3,0);
        \node[OrangeRed] {Tube};
    \end{scope}
    
    \begin{scope}[shift={(11, 0.4)}]
        \draw[-latex, solid, line width=1.7pt, black!80, decorate,
        decoration={snake,amplitude=.4mm,segment length=2mm,post length=1mm}]
        (0,0.5) -- node[midway, above right] {\large $G_{flames}$} (-2.3,3.9);
        \node [tape, left color=Red!60, right color=BurntOrange!60, tape bend bottom=none, tape bend height=8, minimum width=1cm] at (0.2,0.2) {Flames};
        \node [tape, left color=Red, right color=BurntOrange, tape bend bottom=none, tape bend height=7, minimum width=1.5cm] at (0.1,0.1) {Flames};
        \node [tape, left color=Red!50, right color=BurntOrange!50, tape bend bottom=none, minimum width=2cm] at (0,0) {Flames};
    \end{scope}

    \begin{scope}[shift={(-0.6, \p)}]
        \node[black] at (-0.4, 0.7) {Sensor $\lambda$};
        \node[inner sep=0pt] at (0,0) {\includegraphics[width=3cm, angle=180]{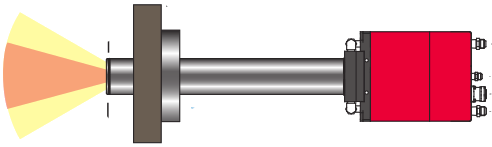}};
    \end{scope}

    \coordinate (A1) at (6, 0);
    \coordinate (B1) at (7.7, \p-1);
    \coordinate (B1P) at (7.3, \p-0.7);
    \coordinate (C1) at (4, \p-0.45);
    \coordinate (D1) at (3.2, \p-0.4);

    \coordinate (A2) at (7, \p);
    \coordinate (B2) at (4, \p);
    \coordinate (C2) at (3.2, \p);

    \coordinate (A3) at (3.2, \p+0.5);  
    
    \coordinate (E1) at (0.8, \p-0.18);
    \coordinate (E2) at (0.8, \p);
    \coordinate (E3) at (0.8, \p+0.18);
    
    \draw[-latex, solid, line width=1.7pt, black!80, decorate, 
    decoration={snake,amplitude=.4mm,segment length=2mm,post length=1mm}] 
    (A1) -- node[midway, above left] {\large $G_{wall}$} (B1); 
    \draw[-latex, solid, line width=1.5pt, black!80, decorate,
    decoration={snake,amplitude=.3mm,segment length=2mm,post length=2mm}] 
    (B1P) -- node[pos=0.4, below] {\large $G_{reflect}$} (C1);  
    \draw[-latex, solid, line width=1pt, black!80, decorate,
    decoration={snake,amplitude=.2mm,segment length=2mm,post length=1mm}]
    (D1) -- node[pos=0.4, below] {$G_{reflect}'$} (E1); 

    \draw[-latex, solid, line width=1.5pt, black!80, decorate,
    decoration={snake,amplitude=.3mm,segment length=2mm,post length=2mm}]
    (A2) -- node[pos=0.4, above] {\large $G_{emit}$} (B2); 
    \path (C2) -- node[pos=0.4] (success) {$G_{emit}'$} (E2);
    \draw[-latex, solid, line width=1pt, black!80, decorate,
    decoration={snake,amplitude=.2mm,segment length=2mm,post length=1mm}] (C2) -- (success) -- (E2); 
    
    \draw[-latex, solid, line width=1pt, black!80, decorate,
    decoration={snake,amplitude=.2mm,segment length=2mm,post length=1mm}] 
    (A3) -- node[pos=0.4, above] {$G_{gas}$} (E3); 


    \node[black, anchor=west] at (8, -0.5) {$G_{in} = G_{wall} + G_{flames} + G_{tube}$};
    \node[black, anchor=west] at (8, -1.0) {$G_{out} = G_{emit} + G_{reflect} + G_{transmit}$};
    \node[black, anchor=west] at (0, -0.5) {$G_{sensor} = G_{gas} + G_{emit}' + G_{reflect}'$};

\end{tikzpicture}
    }
    \caption{Energy exchange by radiation on a simplified furnace tube. Legend: $G$: irradiance $[W/m^2]$, $T$: temperature $[K]$, $\varepsilon$: emissivity, $\lambda$: wavelength $[\mu m]$, $\alpha$: absorption. The subscripts $s$, $w$ and $g$ refer to the tube surface, the furnace wall, and the fuel gas, respectively.
    }
    \label{fig:furnace_diagram}
\end{figure*}

\section{Thermometry Techniques}\label{sec:thermometry}

Temperature measurement in petrochemical industry furnaces is not a straightforward procedure. There are two types of methods for measuring temperature: contact thermometry and non-contact thermometry \cite{saunders, thermometry}. As the name implies, contact thermometers require physical contact with the object of interest before a temperature reading can be obtained. This requirement constrains the spatial sampling to a single point, whereas non-contact methods can be performed remotely and provide dense sampling around the examined area. 

Early approaches to measuring temperature in furnaces relied on direct physical contact. There are many contact thermometers available, e.g., liquid-in-glass thermometers, thermistors, and resistance thermometers. Thermocouples \cite{thermocouples_2} are the most common type of contact thermometer found in industrial applications \cite{thermocouple_1} and are widely used for various tasks throughout the petrochemical industry.
As a result of the thermoelectric effect, this electrical device generates a temperature-dependent voltage, which is then interpreted to determine the temperature.

The applicability of contact sensors is limited due to the complexity of the wiring and instrumentation. 
There are two main issues associated with their use in process tube temperature measurement. The first issue stems from the aggressive chemical nature of a petrochemical furnace's fuel gas environment. Only a few thermocouple types can withstand such aggressive environment, and the thermocouple elements quickly become contaminated, causing a change in their Seebeck coefficient and, as a result, a drift in the measurements \cite{Seebeck1, Seebeck2}. 
The second issue is that proper thermal contact between the thermocouple and the process tube is difficult to achieve, especially considering the high gas flows found in these furnaces. This can result in potentially significant errors in thermocouple readings. Furthermore, there are furnaces whose configuration precludes the use of thermocouples, such as reformer furnaces with tubes located in the center with no access from a sidewall.

Apart from the limitations inherent to the furnace's extreme environment, the thermocouple must be in direct contact with the measuring point, and the corresponding devices must be installed near the tubes attached to the furnace walls, which makes it difficult to pinpoint the exact location of the measuring point. In addition, only the temperature measurement of a specific point in a tube is obtained, which cannot be extrapolated to the entire tube due to temperature non-uniformity.

\newcommand{\G}{$\color{green}\blacktriangle$}
\newcommand{\B}{$\color{red}\blacktriangledown$}
\begin{table}[htbp!]
    \centering
    \resizebox{\linewidth}{!}{
        \begin{tabular}{lll}
        \toprule
        & \textbf{Contact thermometry} & \textbf{Non-contact thermometry} \\ \midrule
        \multirow{6}{*}{\G \textbf{Pros}} 
        & \multirow{6}{*}{\begin{tabular}[c]{@{}l@{}}Sensor availability\\Small sensor size\\Low-cost\\\end{tabular}}
        & Wide temp. range                         \\
        & & High accuracy                            \\
        & & Fast response time                       \\
        & & Transient temp.                          \\
        & & Continuous monitoring                    \\
        & & Interaction-free measurement             \\ \midrule
        \multirow{6}{*}{\B \textbf{Cons}} 
        & Reduced temp. range              & Dependence on object's emissivity        \\
        & Low accuracy                     & Affected by other radiant sources        \\
        & Slow response time               & Measurement in line-of-sight             \\
        & Interferes monitored process     & Disturbed by ambient temp.               \\
        & Influences measured object       & High-cost                                \\
        & Limited by thermal conductivity  & \\
        \bottomrule
        \end{tabular}
    }
    \caption{Characteristic comparison between \textit{average} contact and non-contact infrared thermometry methods.}
    \label{table:comparison2}
\end{table}

As a result, early contact-based approaches were dismissed in favor of \textbf{non-contact measurement} methods (see Table \ref{table:comparison2} for a comparison between contact and non-contact methods). Radiation thermometry, a type of non-contact thermometry, avoids the issues as mentioned earlier \cite{non-contact, sota_ir, Maldague1999}. A radiation thermometer detects infrared radiation emitted by an object without making physical contact with it. Therefore, this instrument can be placed outside the furnace's hostile environment; the only requirement is a clear line of vision to the target tube. The radiation exchange process ensures thermal contact in this case.

When operating from outside the furnace, it is possible to measure different zones with a unique instrument simply by moving it around. As a result of this approach, portable devices such as radiation thermometers \cite{single-point} were developed.
These devices infer temperature from a wavelength range of thermal radiation and can measure the temperature of an industrial furnace using appropriate optical spatial resolution and infrared filtering. Due to its portability and operational simplicity, this method has widely replaced or supplemented thermocouples in industry. 

Despite the performance of these devices, the obtained temperature may be altered and be inaccurate due to peculiar conditions such as distance, target's emissivity, reflected radiance, or furnace gas emission. Thus, radiation measurement is subject to surrounding noise, especially in a harsh and highly radiative environment such as an industrial furnace. 
As a result, while radiation thermometers are easy to use, knowledge of radiation phenomena and material physical properties is required for proper radiation thermometry application \cite{emissivity}.

To overcome the mentioned limitations, \textbf{modulated laser beams} were proposed as a method of measuring the emissivity of the surface as well as the distance to it. This auxiliary sensor helps to extract more information from the surroundings and improves measurement accuracy. However, this method just obtains the temperature of a specific point in the furnace, and the beam cannot measure all tubes across the furnace, considering that it depends on the point of view and the perspective. Besides, the operator must move the device to obtain the temperature of each tube. The sensor's manual operation limits monitoring, making it insufficient to continuously monitor an entire industrial furnace.

In contrast to these approaches, \textbf{novel non-contact thermal mapping} techniques have been proposed for continuous monitoring. These methods employ an imaging radiometer to generate a thermal image of the furnace's interior. 
By measuring the radiation emitted by each point, the operator can obtain an image of the thermal distribution inside the furnace.
Still, the large quantities of combustion gases inside the furnace may distort the measurements and generate noise due to the amount of infrared radiation.
Edwards et al. \cite{gas_wavelength} found that at a wavelength of $3.90 \mu m$, there was a void in the emission spectrum of these combustion gases. Hence, in the presence of hot combustion gases, a narrow band-pass interference filter centered at such wavelength can be used to obtain a precise temperature image of the furnace \cite{saunders}. This thermal mapping approach requires the sensor to use cryogenic cooling to measure the temperature inside the furnace. As an alternative, \textbf{Uncooled focal plane array} (UFPA) sensors were introduced to the market \cite{dual-band, ufpa}.

\section{Error and Uncertainty Analysis}\label{sec:uncertainty_budget}

Although many infrared (IR) thermometry applications do not require knowledge of temperature measurement accuracy, some do, for instance, high radiation environments such as furnaces in the petrochemical industry. The high precision requirements of these furnaces need correction methods to overcome the limitations mentioned above. For such applications, a systematic method should be used to determine its accuracy. 

For temperature measurements, accuracy is the "closeness of agreement between the result of a temperature measurement and a true value of the temperature" \cite{astme344}. 
An accepted method of determining measurement accuracy is to create an error and uncertainty analysis, which provides accurate data for a given measurement.

The analysis presented in this section follows the guideline defined in the Guide to the Expression of Uncertainty in Measurement, commonly referred to as the GUM \cite{kirkup2006introduction}, that establishes general rules for evaluating and expressing uncertainty in measurements. First, the primary sources of uncertainty are identified (\S\ref{sub:sources_error}), and the temperature measurement is modeled with a spectral-band radiation thermometry measurement integral equation (\S\ref{sub:measurement}). Then, a sensitivity analysis is carried out for each model parameter (\S\ref{sub:sensitivity}). Each parameter is successively perturbed from its nominal value, while the remaining parameters stay unchanged at their nominal values, and the difference in the temperature is measured. 
With this procedure, one can analyze how a perturbation (error) in a parameter propagates to the measured temperature. In addition, by repeating the error perturbation for a range of temperatures, one can examine the effect of a distribution of errors, or uncertainty, in an input parameter. 

\subsection{Sources of error}\label{sub:sources_error}

Radiation thermometers have many sources of error that must be considered to establish a high level of confidence in their measurement. A comprehensive overview of these errors applied to a petrochemical furnace is given by Saunders \cite{saunders}, who categorizes the errors into two groups: those associated with the instrument itself and those related to the target and its surroundings (environmental). The most important errors for industrial applications are related to the target and its surroundings. Nevertheless, section \ref{sec:monitoring_system} also addresses errors related to instrument uncertainty, such as instrument stability or temporal variation.

In a petrochemical furnace, there are numerous sources of uncertainties that could be considered when assembling an uncertainty budget for radiation thermometry, e.g., 
target emissivity, reflection (background temperature), absorption and emission (atmospheric losses), spectral response, errors due to flames, scattering errors, size-of-source effect, vignetting, ambient temperature, signal linearization, 
calibration, stability (long term drift), uniformity, noise, readout resolution, among others.

Among these sources of uncertainty, in this article only the most relevant are studied, which are associated with the largest errors \cite{saunders, saunders2008uncertainty, corwin1994temperature, liebmann2008infrared}: 
\begin{enumerate}
    \item \textbf{Emissivity error}: uncertainty in the target's emissivity parameter.
    \item \textbf{Reflection error}: unknown background radiation reflected from the target and detected by the thermometer.
    \item \textbf{Absorption and emission error}: caused by the atmospheric (fuel gas) attenuation of the radiation between the target and the thermometer.
    \item \textbf{Spectral variation}: uncertainty in sensor's wavelength parameter.
\end{enumerate}

Note that in many cases, the effects of atmospheric attenuation and background radiation either are negligible or can be minimized by proper shielding of radiation or choice of wavelengths. 
Nevertheless, in this work, the absorption and emission errors were considered as a relative baseline of the uncertainty introduced in the temperature measurement by these errors.

To quantify the influence of each uncertainty source on temperature accuracy, a measurement model must be defined. In the subsequent section, four different measurement models are introduced, in increasing order of complexity, to account for the errors listed above.

\subsection{Measurement Models}\label{sub:measurement}

The measurement equation used in these error and uncertainty analysis is derived from Planck’s Law. Table \ref{tab:models} summarizes the variables used by each model.

\begin{table}[!htb]
\centering
\resizebox{0.8\linewidth}{!}{
    \begin{tabular}{lccccc}
    \toprule
    Model & $\lambda$ & $\varepsilon$ & $T_w$ & $\alpha$ & $T_g$ \\
    \midrule
    A: Blackbody  & \checkmark  &   &   &   &   \\
    B: Selective Radiator (SR)  & \checkmark & \checkmark &   &   &   \\
    C: SR + Reflections  & \checkmark & \checkmark & \checkmark &   &   \\
    D: SR + Reflections + Gas  & \checkmark & \checkmark & \checkmark & \checkmark & \checkmark \\
    \bottomrule
    \end{tabular}%
}
\caption{Variables included in each model. Legend: $\lambda$: wavelength, $\varepsilon$: emissivity, $T_w$: furnace walls temperature, $\alpha$: gas absorption, $T_g$: fuel gas temperature}
\label{tab:models}
\end{table}

\subsubsection{Model A: Blackbody}

The signal $S$ measured by the monochromatic radiation thermometer is proportional to the blackbody spectral radiance $L_b$ modeled by Planck's Law:

\begin{equation}\label{eq:planck}
L_b(\lambda, T) = \cfrac{c_{1L}}{\lambda^{5}\Big(exp\Big(\cfrac{c_{2}}{\lambda T}\Big)-1\Big)}
\end{equation}

where $\lambda$ refers to the wavelength (usually measured in $\mu m$), $c_{1L}=2hc^2$ is referred as the first radiation constant and $c_2={h c}/{k_B}$ as the second radiation constant. 

\noindent
Note that $h=6.62607015 \, 10^{-34} J s$ is the Planck constant, $k_{B}=1.380649\, 10^{-23} J K^{-1}$ is the Boltzmann constant, and c is the speed of light in the medium ($c_{0}=299792458 m/s$).

One alternative to Planck’s Equation is the Sakuma-Hattori equation \cite{saunders}, which has been suggested as a measurement equation for radiation thermometry error and uncertainty analysis \cite{saunders2008uncertainty}:
\begin{equation}\label{eq:Sakuma-Hattori}
S = \cfrac{C}{exp\Big(\cfrac{c_{2}}{A + B T}\Big)}
\end{equation}

In this paper, the Sakuma-Hattori equation will not be used because of the dynamic nature of the thermometer spectral response and due to furnace's operating temperature above the silver point ($962\degree C$). 

In a spectral-band radiation thermometer, the relationship between the thermometer output signal $S(T)$ and the blackbody temperature $T$ is given by the integral over all wavelengths of the product of the spectral power imaged onto the detector and its spectral responsivity.

\begin{equation}\label{eq:model1A}
S(T) = \int_0^\infty R(\lambda) \cdot L_b(\lambda, T) \: d\lambda
\end{equation}

where $R(\lambda)$ is the thermometer's absolute spectral responsivity. In case the spectral responsivity is limited to a range, the integral is defined over the interval $[\lambda_1,\lambda_2]$:

\begin{equation}\label{eq:model1B}
S(T) = \int_{\lambda_1}^{\lambda_2} R(\lambda) \cdot L_b(\lambda, T_s) \: d\lambda
\end{equation}

Equation \ref{eq:model1B} defines measurement model A, and it serves as the foundation for all three subsequent models. This model assumes an isolated blackbody tube with surface temperature $T_s$, with no reflections or interactions with other objects. Inferring the temperature of the target surface $T_s$ using equation \ref{eq:model1B} requires solving an unconstrained optimization problem.

\subsubsection{Model B: Selective Radiator}


Model B is an extension on Model A that considers an isolated, opaque, diffuse, and selective radiator body with emissivity $\varepsilon(\lambda)$, which is a function of the wavelength $\lambda$.

\begin{equation}\label{eq:model2}
S(T) = \int_{\lambda_1}^{\lambda_2} \varepsilon(\lambda) \cdot R(\lambda) \cdot L_b(\lambda, T_s) \: d\lambda
\end{equation}

\subsubsection{Model C: Selective Radiator + Reflections}

In a steam reformer, the tubes are usually the coldest objects, with the walls, floor, and ceiling measuring as much as $300$ to $400 \degree C$ higher. Therefore, errors due to reflected radiation make the thermometer report a higher temperature than the actual value.

Model C extends Model B and takes into account surrounding bodies and their temperature in the measurement equation. In this case, the infrared signal received by the thermometer from the tube is the summation of the radiation emitted by the tube, and the radiation reflected off the tube that originated from other surrounding objects (furnace walls): $G_{sensor} = G_{emit} + G_{reflect}$.

\begin{equation}\label{eq:model3}
\begin{split}
S(T) = & \int_{\lambda_1}^{\lambda_2} \varepsilon(\lambda) \cdot R(\lambda)  \cdot L_b(\lambda, T_s) \: d\lambda \: + \\
& \int_{\lambda_1}^{\lambda_2} (1-\varepsilon(\lambda)) \cdot R(\lambda) \cdot L_b(\lambda, T_w) \: d\lambda
\end{split}
\end{equation}

In equation \ref{eq:model3}, the signal $S$ for the measured temperature $T$ depends on the effective background temperature $T_{w}$, the true temperature of the tube being measured ($T_{s}$) and its emissivity $\varepsilon$. Bodies are assumed to be opaque, diffuse, and selective radiators. 
By the energy balance at the body surface, light must be absorbed, reflected or transmitted: $\varepsilon + \rho + \tau = 1$, where $\rho$ is the reflectance and $\tau$ the transmittance. Thus, considering that the transmittance $\tau$ of an opaque body in equilibrium is zero, the emissivity and the reflectance are complementary ($\rho + \varepsilon = 1$).
Note that in this model, the emissivity is not considered a function of the incidence angle. The effective background temperature $T_w$ is equal to a weighted average of the temperatures of all of the surrounding objects that can be seen by the tube. 

\subsubsection{Model D: Selective Radiator + Reflections + Gas Absorption}

\begin{figure}[!htb]
    \centering
    \resizebox{1.0\linewidth}{!}{
    \begin{tikzpicture}[shift={(0,0)}]
    \pgfmathsetmacro{\n}{10}
    \pgfmathsetmacro{\m}{8.25}
    \pgfmathsetmacro{\p}{5}
    \pgfmathsetmacro{\h}{0.5}
    \pgfmathsetmacro{\s}{1.35}


    \draw[Sepia, ultra thick, double, line cap=rect] (0,0) -- (\n, 0);
    \draw[Sepia, ultra thick, double, line cap=rect] (\n, 0) -- (\n, \m);
    \draw[Sepia, ultra thick, double, line cap=rect] (\n, \m) -- (0, \m);
    \draw[Sepia, ultra thick, double, line cap=rect] (0,0) -- (0, \p - \h);
    \draw[Sepia, ultra thick, double, line cap=rect] (0,\m) -- (0, \p + \h);
    
    \begin{scope}[shift={(3, 0)}]
        \draw[Sepia] (0, 0) to[out=45,in=180] (0.7,1) node[anchor=west, scale=\s] {$T_w$};
    \end{scope}
    \node[Sepia, anchor=south west] at (0.2,0.2) {Furnace Walls};

    \begin{scope}[shift={(4, \p)}]
        \node [cloud, draw, cloud puffs=15, cloud puff arc=140, 
        aspect=1.5, rotate=90, inner ysep=0.7em, scale=2, fill=white, thick] at (0,0) {};
        \draw[black] (0.3, 1) to[out=45,in=0] (0,1.8) node[anchor=base east, scale=\s] {$T_g, \alpha(\lambda)$};
        \node[black] at (0, 0.5) {Gas};
    \end{scope}

    \begin{scope}[shift={(8, \p)}]
        \filldraw[color=OrangeRed, fill=OrangeRed!5, very thick] (0,0) circle (1);
        \draw[OrangeRed] (-0.7, 0.7) to[out=135,in=180] (-0.4,1.8) node[anchor=west, scale=\s] {$T_s, \varepsilon(\lambda)$};
        \node[OrangeRed] {Tube};
    \end{scope}

    \begin{scope}[shift={(-0.6, \p)}]
        \node[black] at (-0.4, 0.7) {Sensor $\lambda$};
        \node[inner sep=0pt] at (0,0) {\includegraphics[width=3cm, angle=180]{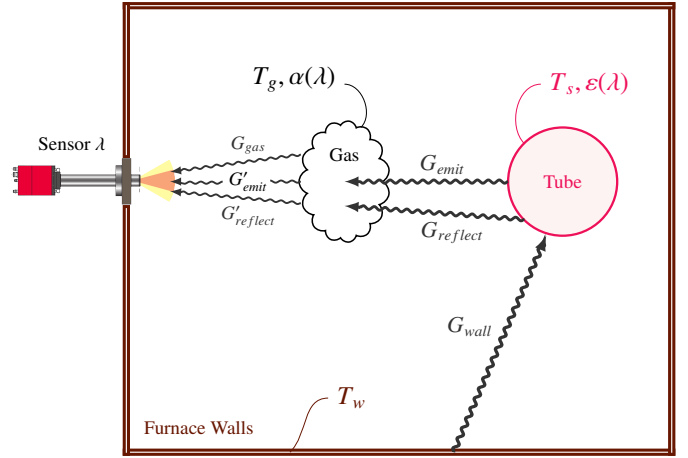}};
    \end{scope}

    \coordinate (A1) at (6, 0);
    \coordinate (B1) at (7.7, \p-1);
    \coordinate (B1P) at (7.3, \p-0.7);
    \coordinate (C1) at (4, \p-0.45);
    \coordinate (D1) at (3.2, \p-0.4);

    \coordinate (A2) at (7, \p);
    \coordinate (B2) at (4, \p);
    \coordinate (C2) at (3.2, \p);

    \coordinate (A3) at (3.2, \p+0.5);  
    
    \coordinate (E1) at (0.8, \p-0.18);
    \coordinate (E2) at (0.8, \p);
    \coordinate (E3) at (0.8, \p+0.18);
    
    \draw[-latex, solid, line width=1.7pt, black!80, decorate, 
    decoration={snake,amplitude=.4mm,segment length=2mm,post length=1mm}] 
    (A1) -- node[midway, above left] {\large $G_{wall}$} (B1); 
    \draw[-latex, solid, line width=1.5pt, black!80, decorate,
    decoration={snake,amplitude=.3mm,segment length=2mm,post length=2mm}] 
    (B1P) -- node[pos=0.4, below] {\large $G_{reflect}$} (C1);  
    \draw[-latex, solid, line width=1pt, black!80, decorate,
    decoration={snake,amplitude=.2mm,segment length=2mm,post length=1mm}]
    (D1) -- node[pos=0.4, below] {$G_{reflect}'$} (E1); 

    \draw[-latex, solid, line width=1.5pt, black!80, decorate,
    decoration={snake,amplitude=.3mm,segment length=2mm,post length=2mm}]
    (A2) -- node[pos=0.4, above] {\large $G_{emit}$} (B2); 
    \path (C2) -- node[pos=0.4] (success) {$G_{emit}'$} (E2);
    \draw[-latex, solid, line width=1pt, black!80, decorate,
    decoration={snake,amplitude=.2mm,segment length=2mm,post length=1mm}] (C2) -- (success) -- (E2); 
    
    \draw[-latex, solid, line width=1pt, black!80, decorate,
    decoration={snake,amplitude=.2mm,segment length=2mm,post length=1mm}] 
    (A3) -- node[pos=0.4, above] {$G_{gas}$} (E3); 

\end{tikzpicture}
    }
    \caption{Geometry of the furnace and radiation sources included in model D.}
    \label{fig:furnace_diagram_model}
\end{figure}

Fig. \ref{fig:furnace_diagram_model} shows a simplified geometry of a refinery furnace, consisting of a unique tube at surface temperature $T_s$ and with surface emissivity $\varepsilon$ with uniform wall temperature $T_w$. The atmosphere inside the furnace is modelled with a gas at uniform temperature $T_g$, absorption coefficient $\alpha$ and a distance to the target surface $l$. Note that both emissivity and the absorption coefficient are a function of the wavelength $\lambda$: $\varepsilon(\lambda)$, $\alpha(\lambda)$. 

\begin{equation}\label{eq:model4}
\begin{split}
S(T) = 
& \int_{\lambda_1}^{\lambda_2} (1-l \cdot \alpha(\lambda))  \cdot \varepsilon(\lambda)  \cdot  R(\lambda)  \cdot L_b(\lambda, T_s) \: d\lambda \: + \\
& \int_{\lambda_1}^{\lambda_2} (1-l \cdot \alpha(\lambda)) \cdot (1-\varepsilon(\lambda)) \cdot R(\lambda) \cdot L_b(\lambda, T_w) \: d\lambda \: + \\
& \int_{\lambda_1}^{\lambda_2} \alpha(\lambda) \cdot R(\lambda) \cdot L_b(\lambda, T_g) \: d\lambda \: \\
\end{split}
\end{equation}

Under this assumption, the infrared signal received by the thermometer from the tube wall is a summation of the radiation emitted by the tube and the radiation reflected off the tube that originated from other surrounding objects. This summation is then attenuated by the gases present in the furnace's atmosphere. Since the thermometer cannot differentiate between these sources, the resulting temperature measurement has errors that depend primarily on the ratio of emitted radiation to reflected radiation.
The irradiance $G$ contributions represented in Fig. \ref{fig:furnace_diagram_model} can be defined as:

\begin{align}
G_{wall} &\propto L_b(\lambda, T_w) \\
G_{reflect} &\propto (1-\varepsilon(\lambda)) \cdot G_{wall} \\
G_{reflect}' &\propto (1-l \cdot \alpha(\lambda)) \cdot G_{reflect} \\
G_{emit} &\propto (1-\varepsilon(\lambda)) \cdot L_b(\lambda, T_s) \\
G_{emit}' &\propto (1-l \cdot \alpha(\lambda)) \cdot G_{emit} \\
G_{gas} &\propto l \cdot \alpha(\lambda) \cdot L_b(\lambda, T_g) \\
G_{sensor} &\propto G_{emit}' + G_{reflect}' + G_{gas}
\end{align}

\subsection{Sensitivity Analysis}\label{sub:sensitivity}

In this section, a parameter sensitivity analysis is conducted for measurement models B, C, and D. Sensitivity analysis is the study of how the uncertainty in the output of a model can be apportioned to different sources of uncertainty in the model input \cite{saltelli2002sensitivity}.

Table \ref{table:sensitivity} shows the parameters for the sensitivity analysis, with their nominal value, range, and units of measure. The nominal and range values are chosen based on a typical steam reformer's operating conditions. It should be noted that not all parameters apply to all the models (see Table \ref{tab:models}). The sensitivity analysis is performed in the following manner: each parameter is successively perturbed from its nominal value, while the remaining parameters stay unchanged at their nominal values. Then, the difference in the tube temperature $\Delta T_{nom}$ is measured. This procedure is also repeated for a variety of tube temperature values to examine its effect on the measured difference $\Delta T_{nom}$. Graphical results of this analysis are included on \ref{app:sensitivity}. 

\begin{table}[htbp!]
    \centering
    \resizebox{1\linewidth}{!}{%
    \begin{tabular}{lllll}
        \toprule
        Parameter & Description & Nominal value & Range & Unit \\ \toprule
        $\lambda$ & Wavelength & 3.95 & 3.7 - 4.2 & $\mu m$ \\
        $\varepsilon$ & Emissivity & 0.82 & 0.72 - 0.92 & - \\
        $\alpha$ & Absorption & 0.05 & 0.0 - 0.1 & - \\
        $T_{w}$ & Wall temperature & 1105 & 1030 - 1180 & $\degree C$ \\
        $T_{g}$ & Gas temperature  & 980 & 880 - 1080 & $\degree C$ \\
        $T_{s}$ & Tube temperature & 950 & 880 - 1030 & $\degree C$ \\
        \bottomrule
    \end{tabular}
    }
    \caption{Parameter sensitivity analysis: nominal value, range and units.}
    \label{table:sensitivity}
\end{table}

Figs. \ref{fig:sensitivityM2}, \ref{fig:sensitivityM3}, and \ref{fig:sensitivityM4} depict the results of the sensitivity analysis for models B, C and D respectively. Regarding model B, Fig. \ref{fig:sensitivityM2A} illustrates the influence of the wavelength $\lambda$ on the measured tube temperature $T_s$. A dotted vertical line is drawn at the nominal value, in this case, 3.95 $\mu m$. The perturbed parameter is located on the x-axis, and the difference in the measured tube temperature $\Delta T_{nom}$ is plotted on the y-axis. Each line is associated with a nominal tube temperature value. In this case, Fig. \ref{fig:sensitivityM2A} shows that there is a positive correlation between a change in the wavelength parameter and the tube temperature, but the effect of that change is small (between -5 and 5 $\degree C$) for the operating temperature of the furnace. 

Continuing with model B, Fig. \ref{fig:sensitivityM2B} shows a strong negative correlation between a perturbation in the emissivity parameter and the tube temperature that varies heavily (between -50 and 60 $\degree C$). Therefore, the emissivity parameter has a greater influence than the wavelength parameter ($\sim$ x10 ratio), based on measurement model B. In order to facilitate the comparison of the influence of the model parameters, Fig. \ref{fig:uncertaintyM2} shows a different perspective of this statement. In this case, the nominal tube temperature $T_s$ is plotted on the x-axis and the uncertainty $u_{nom}$ of the y-axis. The uncertainty $u_{nom}$ is defined as the maximum absolute difference in temperature $\Delta T_{nom}$ when a parameter is perturbed from its nominal value. 
Apart from the uncertainty introduced by each parameter, the combined uncertainty $u_{c}$ has also been included. Uncertainties add in quadrature $u_{c} = \sqrt{\sum_{i} u_{i}^2}$.
The expanded uncertainty $U$ can be calculated as the product of the combined uncertainty $u_{c}$ and the coverage factor $k$: $U=k \cdot u_{c} = k \cdot \sqrt{\sum_{i} u_{i}^2}$. A $95\%$ confidence interval is established with a coverage factor of $k=1.96$. Expanded uncertainty is excluded from figures \ref{fig:uncertaintyM2}, \ref{fig:uncertaintyM3} and \ref{fig:uncertaintyM4} due to its disproportionate scale.

In the case of Model B (see Fig. \ref{fig:uncertaintyM2}) the total uncertainty is almost entirely due to the emissivity parameter. 
Similar analyses have been carried out for measurement models C and D. Concerning model C there is a slight positive correlation with the wavelength $\lambda$, a strongly positive one with the emissivity $\varepsilon$ and a strong negative correlation with the wall temperature $T_w$ (see Fig. \ref{fig:sensitivityM3}). Note how the difference in temperature $\Delta T_{nom}$ due to the emissivity $\varepsilon$ approaches zero as the tube temperature $T_s$ and wall temperature $T_w$ get closer (see Fig. \ref{fig:uncertaintyM3}). According to model C, the emissivity is neutralized when both the tubes and the walls have the same temperature. For measurement model D, a comparative analysis can be performed (see Fig. \ref{fig:sensitivityM4}): both the wavelength $\lambda$, the fuel gas temperature $T_g$ and its absorption coefficient $\alpha$ have a small influence ($<10\degree C$) on the measured tube temperature, relative to the effect of the emissivity $\varepsilon$ and the wall temperature $T_w$ ($\sim$ 20-40$\degree C$). These results support the findings of other authors \cite{saunders, saunders2008uncertainty, willmott2016potential, liebmann2008infrared}, who concluded that the effects of atmospheric attenuation are negligible.

In this section, a parameter sensitivity analysis has been performed for multiple measurement models, yielding the impact of each model parameter on the output temperature. Based on the selected model, refinery furnace operators can be informed about the uncertainty in the measured temperature. It should be noted that this procedure can be extended to other, more complex thermometry models.

\section{Monitoring System}\label{sec:monitoring_system}

In this section, the system for continuous monitoring of industrial furnaces is introduced. The measurement correction methodology is integrated into the thermal imagery acquisition system, designed with an end-to-end computing architecture, and handles everything from thermal imagery acquisition to analysis.

\subsection{Thermal imagery acquisition}
\label{sub:thermalimagery}

In this work, a camera with an amorphous silicon microbolometer focal plane array (FPA) infrared sensor was used. FPA sensors are widely used \cite{fpa_1, fpa_2, fpa_3} and have proven to be a significant advance in radiometric imagery in hot atmospheres. Performance and cost factors were considered to decide where to place the camera in the furnace. Installing a camera inside industrial furnaces is both costly in terms of maintenance and complicated in terms of operation and installation. 

\begin{figure}[!htb]
    \centering
    \includegraphics[width=\linewidth]{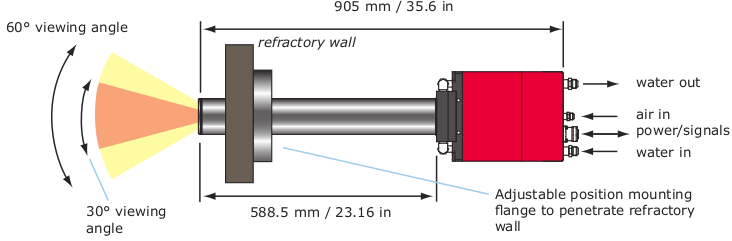}
    \caption{ FTI-Eb Borescope thermal imaging camera's diagram \cite{land}.}
    \label{fig:camera}
\end{figure}

To overcome such limitations, we used a radiometric infrared borescope camera \cite{land}. A borescope is an optical instrument used to inspect narrow and difficult-to-reach cavity areas (see Fig. \ref{fig:camera}). This instrument allows the infrared camera to be placed outside the furnace and introduces the borescope to acquire the image. In order to ensure equipment safety and correct operation performance during the continuous monitoring, a water-cooled rugged housing keeps it at operating temperature. Furthermore, the camera employs non-uniformity correction (NUC) to compensate for minor detector drift caused by the instrument's heating.

The borescope can be inserted through mounting openings in the furnace walls. These holes are coated with a mounting standpipe. This high-temperature-resistant tube allows the camera to be inserted in the right direction and position, ensuring consistency between different captures. The camera was also mounted and anchored to a flange attached to the standpipe. In this manner, the camera position is guaranteed to be constant, even when temporarily removed.

The mounting standpipe does not penetrate the wall, preventing overheating and unnecessary exposure to the environment. The borescope exceeds the extension of the standpipe at its end, avoiding interference of the standpipe during the image capture, i.e., by reflections. Furthermore, the standpipe being shorter than the wall width allows sealing the hole using a ceramic plug. In this manner, when the camera is not in operation, the furnace does not suffer any heat losses. 

\subsection{Surrogate model}

Spectral radiation thermometers typically include a look-up table built into the data acquisition module. These look-up tables are limited to elementary models, such as the inverse of Planck's law, which assumes a blackbody object and ignores major environmental errors. A potential alternative to look-up tables is surrogate models, which are statistical approximations of high-fidelity models.

Combining both flexibility and accuracy, we propose \textit{deep learning}-based surrogate models for fast inference in radiation thermometry error-correction models. Measurement model D ($M_D$, from equation \ref{eq:model4}) is selected as the underlying model, as it considers multiple environmental errors. 
Model $M_D$ estimates the temperature of a tube at temperature $T_s$, measured by a sensor with spectral responsivity $R(\lambda)$, given wall temperature $T_w$, fuel gas temperature $T_g$ and properties $\varepsilon(\lambda)$ and $\alpha(\lambda)$. Each parameter is defined over a range of values, as summarized on Table \ref{table:parameters}. 
Note that the functional form of both $\varepsilon(\lambda)$ and $\alpha(\lambda)$ follow a bell-shaped function parametrized by its mean value $\mu$, deviation $\sigma$ and height or maximum value $h$: $f(\lambda) = h \cdot exp(-(\lambda - \mu)^{2} / (2\sigma^2))$, as illustrated on Fig. \ref{fig:gaussian} and Table \ref{table:parameters}.
Depending on the scenario, parameter search ranges can be freely modified, and models can be retrained and adapted to new domains.

\pgfmathdeclarefunction{gauss}{3}{%
  \pgfmathparse{#3*exp(-((x-#1)^2)/(2*#2^2))}%
}

\begin{figure}[!htb]
    \centering
    \resizebox{1.0\linewidth}{!}{
    \begin{tikzpicture}
    \begin{axis}[
        domain=3.3:4.6, 
        xmin=3.3,
        xmax=4.6,
        ymin=0,
        ymax=1.0,
        samples=50,smooth,
        enlargelimits=true,
        xlabel = $\lambda \: {[\mu m]}$,
        ylabel = $\varepsilon(\lambda)$,
        xmajorgrids=true,
        ymajorgrids=true,
        yminorgrids=true,
        grid style=dotted
    ]
        \addplot[black] coordinates {(3,0) (5,0)};
        \addplot[black, dashed] coordinates {(3.9,0.82) (5,0.82)};
        \addplot[black, dashed] coordinates {(3.9,0) (3.9,0.82)};
        \addplot[black, latex-latex] coordinates {(3.6,0.27) (4.2,0.27)} node[midway, fill=white] {$\sigma_{\varepsilon}$};
        \addplot[teal, thick] {gauss(3.9,0.2,0.82)};
        \addplot[black, latex-latex] coordinates {(4.6,0.0) (4.6,0.82)} node[midway, fill=white] {$h_{\varepsilon}$};
        \node[below] at (axis cs:3.9,0) {$\mu_{\varepsilon}$};
    \end{axis}
    \end{tikzpicture}
    \begin{tikzpicture}
    \begin{axis}[
        domain=3.3:4.6, 
        xmin=3.3,
        xmax=4.6,
        ymin=0,
        ymax=0.2,
        samples=50,smooth,
        enlargelimits=true,
        xlabel = $\lambda \: {[\mu m]}$,
        ylabel = $\alpha(\lambda)$,
        xmajorgrids=true,
        ymajorgrids=true,
        yminorgrids=true,
        grid style=dotted,
        yticklabels={$0$, $0$, $0.05$, $0.1$, $0.15$, $0.2$}
    ]
        \addplot[black] coordinates {(3,0) (5,0)};
        \addplot[black, dashed] coordinates {(3.9,0.15) (5,0.15)};
        \addplot[black, dashed] coordinates {(3.9,0) (3.9,0.15)};
        \addplot[black, latex-latex] coordinates {(3.6,0.0455) (4.2,0.0455)} node[midway, fill=white] {$\sigma_{\alpha}$};
        \addplot[magenta, thick] {gauss(3.9,0.2,0.15)};
        \addplot[black, latex-latex] coordinates {(4.6,0.0) (4.6,0.15)} node[midway, fill=white] {$h_{\alpha}$};
        \node[below] at (axis cs:3.9,0) {$\mu_{\alpha}$};
    \end{axis}
    \end{tikzpicture}
    }
    \vspace{-1em}
    \caption{Functional form of the emissivity $\varepsilon(\lambda)$ and the gas absorption $\alpha(\lambda)$, parametrized by the mean value $\mu$, deviation $\sigma$ and height or maximum value $h$.}
    \label{fig:gaussian}
\end{figure}
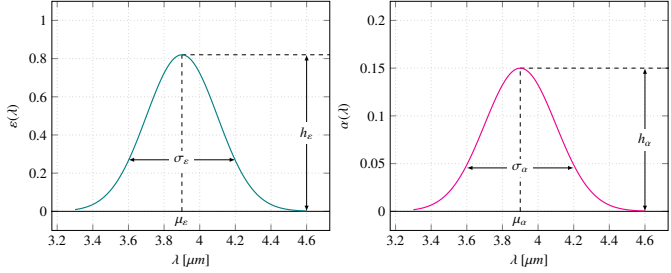
\begin{table}[htbp!]
    \centering
    \resizebox{1\linewidth}{!}{%
    \begin{tabular}{llll}
        \toprule
        Parameter & Description & Range & Unit\\ \toprule
        $T_{s}$ & Tube temperature & 800 - 1200 & $\degree C$ \\
        $T_{w}$ & Wall temperature & 800 - 1300 & $\degree C$ \\
        $T_{g}$ & Gas temperature  & 500 - 1000 & $\degree C$ \\
        $h_{\varepsilon}$ & Emissivity height  & 0.65 - 0.95  & $-$ \\
        $\mu_{\varepsilon}$ & Emissivity's wavelength mean  & 3.3 - 4.6 & $\mu m$ \\
        $\sigma_{\varepsilon}$ & Emissivity's wavelength deviation & 0.2 - 1.8 & $\mu m$  \\
        $h_{\alpha}$ & Gas absorption height  & 0.0 - 0.2 & $-$  \\
        $\mu_{\alpha}$ & Gas absorption's wavelength mean  & 3.3 - 4.6 & $\mu m$  \\
        $\sigma_{\alpha}$ & Gas absorption's wavelength deviation  & 0.2 - 1.8 & $\mu m$\\
        \bottomrule
    \end{tabular}
    }
    \caption{Model D input variables description, search range and units.}
    \label{table:parameters}
\end{table}
\begin{figure}[!htb]
    \centering
    \resizebox{0.9\linewidth}{!}{
\begin{tikzpicture}

    \node at (0,0) [anchor=west, line width=1pt, minimum height=1cm, text width=3cm, draw, align=center, rounded corners=.2cm, Violet] {Measurement Model};
    \node at (0,-2) [anchor=west, line width=1pt, minimum height=1cm, text width=3cm, draw, align=center, rounded corners=.2cm, BrickRed] {Surrogate Model};

    \draw[-latex, solid, line width=1pt, black!60]
    (-0.75,0.5) -- node[pos=0, fill=white, left, thin, rounded corners=.1cm, align=center, text width=1.6cm] 
    {$T_w$, $T_g$ $\varepsilon(\lambda)$, $\alpha(\lambda)$} (0,0.1);

    \draw[-latex, solid, line width=1pt, black!60]
    (4,-2.5) -- node[pos=0, fill=white, right, thin, rounded corners=.1cm, align=center, text width=1.6cm] 
    {$T_w$, $T_g$ $\varepsilon(\lambda)$, $\alpha(\lambda)$} (3.25,-2.1);

    \node at (-1.25,-1) [draw, fill=white, left, thick, rounded corners=.1cm] {\large $T_s$} ;
    \node at (4.5, -1) [draw, fill=white, right, thick, rounded corners=.1cm] {\large $S$} ;

    \draw[-latex, solid, line width=1pt, Violet] (-1.25,-1) --  (0,-0.1);
    \draw[-latex, solid, line width=1pt, Violet] (3.25,0) --  (4.5,-1);
    
    \draw[latex-, solid, line width=1pt, BrickRed] (-1.25,-1) --  (0,-2);
    \draw[latex-, solid, line width=1pt, BrickRed] (3.25,-1.9) --  (4.5,-1);

\end{tikzpicture}
    }
    \caption{Measurement and surrogate model: input and output parameters. Legend: $\lambda$: wavelength, $\varepsilon$: emissivity, $\alpha$: gas absorption, $T_w$: furnace walls temperature, $T_g$: fuel gas temperature, $T_s$: tube temperature, $S$: measured radiation }
    \label{fig:surrogate}
\end{figure}
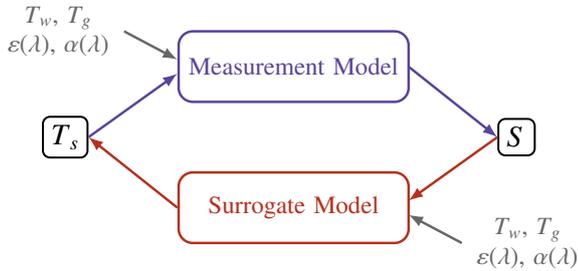

The surrogate model has to learn the inverse function, that is, estimating the tube temperature $T_s$ (output) from the measured radiation (input), given the parameters $T_w$, $T_g$, $\varepsilon(\lambda)$, and $\alpha(\lambda)$ (see Fig. \ref{fig:surrogate}). Due to the high accuracy in complex regression tasks, \textit{Deep Neural Networks} (DNN) are proposed. In this regard, to choose the most effective DNN topologies, an experiment is conducted in which 3500 topologies are created. Among all these topologies, we have selected the best-performing model.

\begin{table}[htbp!]
    \centering
    \resizebox{0.65\linewidth}{!}{
    \begin{tabular}{llll}
        \toprule
        Layer & Neurons & Activation & {\# Parameters} \\ \toprule
        Input & 9 & - & - \\
        Dense & 96 & ReLu \cite{relu} & 864 \\
        Dense & 125 & ReLu & 12000 \\
        Output & 1 & Linear & 125\\
        \midrule
        Total & 231 & - & 12989 \\
        \bottomrule
    \end{tabular}
    }
    \caption{Surrogate deep neural network model's topology.}
    \label{table:final_model}
\end{table}

The model's topology is shown in Table \ref{table:final_model}. 3500 random DNN topologies were evaluated in order to select the most effective combination of layers and neurons.
The neural network layers are fully connected and without bias.
The model was trained for 200 epochs using the Adam optimization algorithm \cite{adam}, with a learning rate of $0.001$ and \textit{Mean Square Error} (RMS$_{error}$) as loss function.
After the training, the model was validated with one million data samples. Experiments were run on an Intel i5-9400F processor with six threads at 2.90 GHz and 16 GB of RAM.
The model performed accurately, obtaining an RMS$_{error}$ of $0.1511 \degree C$ and an inference time of $42.19$ ms. On the contrary, solving the optimization problem with the bisection method (scipy) took $324.3$ ms.

Considering the obtained results, it can be stated that DNN models can be used as a surrogate model of the measurement model $M_D$ defined on section \ref{sec:uncertainty_budget}. In comparison with the underlying physical model $M_D$, the inference speed improvement is nearly x10. This enables efficient measurement correction of radiation thermometry imagery in a variety of industrial furnaces. The proposed DNN topology's lightness makes it easier to retrain models and tune the parameters to improve performance in other conditions (e.g., spectral responsivity of the sensor, gas temperature range). 

\begin{figure*}[!htb]
    \centering
    \includegraphics[width=1.0\linewidth]{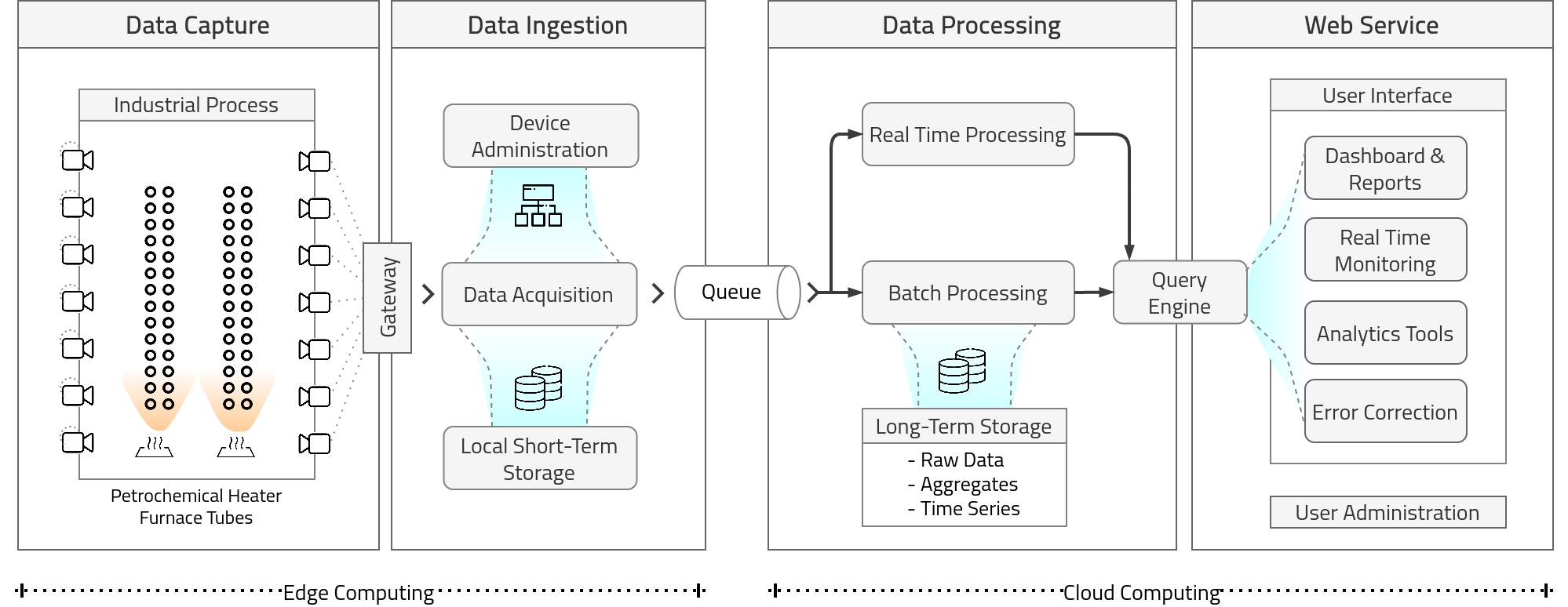}
    \caption{Monitoring system architecture. The edge-computing unit \textit{(left side)} is composed of \textbf{the image acquisition process} which is performed in the furnace, and \textbf{the image management process}, performed in the plant's data center. The cloud-computing unit \textit{(right side)}, is composed of \textbf{the data processing process} which is performed in a cloud server, and \textbf{data analytic platform}, accessible for the domain expert.}
    \label{fig:architecture}
\end{figure*}
\begin{figure}[!htb]
    \centering
    \includegraphics[width=1\linewidth]{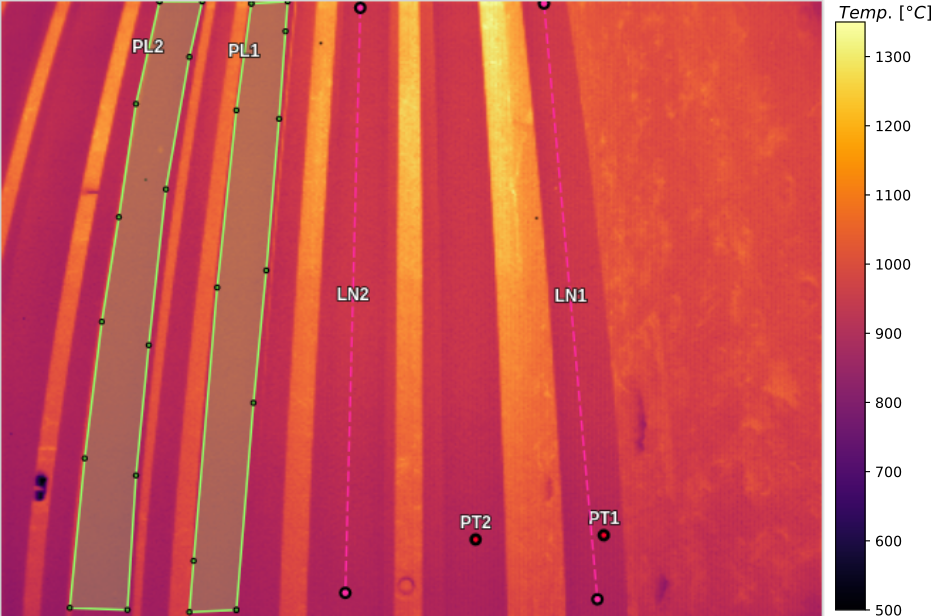}
    \caption{ Thermal image of the furnace: selection of different regions of interest on the tubes, using points, lines and polygons.}
    \label{fig:platform_2}
\end{figure}
\begin{figure}[!htb]
    \centering
    \includegraphics[width=0.8\linewidth]{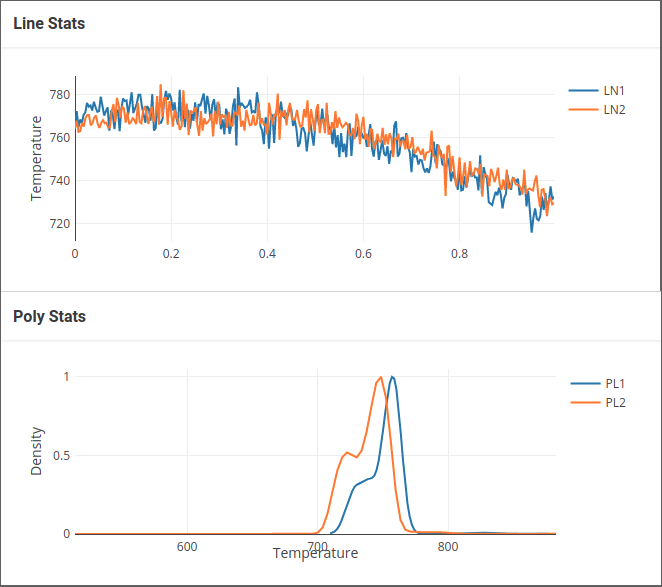}
    \caption{ Selection tool: visualization of the measurements made in the geometric areas selected by the user at Fig. \ref{fig:platform_2}.}
    \label{fig:platform_3}
\end{figure}

Furthermore, by using this technique the expert user is able to change model's inputs in real time, providing accurate emissivity \cite{de2020updated} or gas absorption data for each area or point of the furnace. Thus, the accuracy of the model is complemented by its adaptability to new data.

\subsection{Architecture}

The thermal imaging procedure requires an architecture to coordinate data acquisition, management, and exploitation. In this section we present a computing architecture for integrating the presented monitoring system.

The proposed architecture is able to manage several thermal cameras located throughout an industrial plant. Performing multiple data acquisition processes and analyzing the data in the same computational unit can cause infrastructure congestion. As a result, the architecture was separated into edge- and cloud-computing units, as illustrated in Fig. \ref{fig:architecture}.

Data is collected and processed at the edge computing unit before being ingested into the cloud-computing platform. Data acquisition is performed in an industrial furnace, using above stated methodology. In this regard, a custom software tool was designed to improve the camera's acquisition procedure and enhance the data quality of the thermal images. The tool serves as an orchestration system, controlling the multiple cameras installed in the plant. In this sense, every camera has its corresponding instance in the capture process that is triggered simultaneously. Technically image capture is performed asynchronously, which does not affect the measurement accuracy given that the capture delay is less than half a second and the thermal inertia in an industrial furnace is very low.

The cameras are connected to the plant's data center via a gateway. This data center manages the camera network and orchestrates data acquisition in each furnace at the same time. The collected data is then managed by a local short-term storage system before being uploaded to the cloud platform via a queuing system. 

After the data is uploaded to the cloud-computing unit, it is processed in order to be accessible to the domain expert users. At this stage, a batch processing service generates and extracts information from the uploaded data. In this service, data is processed to obtain time-series instances and aggregated thermal mapping data. Subsequently, this data is stored in a long-term storage service, which can be accessed and managed via the web platform. At last, data serving is handled by a query engine, which optimizes data requests in order to manage computing resources efficiently. For the streaming functionality, a real-time data serving service is provided.

Finally, in the web platform, the domain expert user will access the analytical services, in order to explore and exploit the acquired data. 
This service includes the previously mentioned surrogate models, which correct the data in real time based on the emissivity $\varepsilon$ and gas absorption $\alpha$ inputs provided by the domain expert. These inputs can be inserted for each point of the image, using a mask to assign values. 

In addition, the web platform includes an interactive dashboard with analytical tools focusing on time series and machine learning techniques. With these tools, domain experts can detect furnace anomalies and generate an automated report. 

\subsection{Real scenario integration}\label{sec:plant_integration}

The architecture presented here has been integrated into the infrastructure of a petrochemical refinery at various stages of the refinery process, with cameras installed in several units. With that, the functioning of the system has been validated in a real scenario.

The selection tool, one of many tools available in the web interface, is now introduced as an example. Domain experts can use the selection tool to select regions of interest in the tubes by adding points, lines, and polygons (see Fig. \ref{fig:platform_2}). 
The application then displays a plot with selected geometry areas' measurements (see Fig. \ref{fig:platform_3} for an example of measurements along two lines and inside two polygons defined by the user). Depending on the type of geometry the plot represents different information: a) \textit{Point}: mean value and standard deviation within one-pixel of the selected point. b) \textit{Line}: ordered value at every pixel that intersects with the drawn line. c) \textit{Polygon}: the distribution of measurement values within the drawn polygonal area.
This information is useful to analyze the evolution of the performance and temperature distribution in the furnace. Using this tool, the expert can monitor what is happening inside the furnace and make decisions to avoid potential failures or power outages.

\section{Conclusions}\label{sec:conclusions}

In this paper, a methodology for budgeting the major sources of error and uncertainty during temperature measurements in a petrochemical furnace scenario has been presented. The contribution of each main source of error has been quantified for three different measurement models. This procedure, however, can be extended to more complex mathematical models that include other environmental effects. Based on the selected model, refinery furnace operators can be informed about the uncertainty in the measured temperature, which is a valuable information for improving furnace control and monitoring. 

The measurement model has been integrated into a complete solution for continuous monitoring of industrial furnaces. This solution manages the acquisition of thermal imagery, using multiple borescope cameras introduced into the furnace and anchored to the wall structure via of flanges and standpipes. In addition, using an end-to-end computing architecture the system is capable to orchestrate data acquisition, processing and analysis. In this sense, using light-weight deep neural network, the systems replicates the complex radiation thermometry measurement model with high precision, yielding an inference speed improvement of nearly x10. This enables efficient and accurate measurement of radiation thermometry imagery in a variety of petrochemical industry furnaces. In addition, the accuracy of the model is enhanced by its adaptability to new data, as expert users can change the model's inputs in real time. Furthermore, domain expert users are able to monitor the furnace's operation using user-friendly interaction interfaces that provide analytical tools to investigate and analyze the performance.

A real-world application case in a petrochemical plant was presented to validate the proposed system's functionality, demonstrating that the proposed monitoring strategy is more than just a theoretical-conceptual framework. The system's deployment in a real-world petrochemical plant has contributed to a better understanding of the approach's potential limitations and installation challenges.

Overall, the proposed solution demonstrates the viability of precise industrial furnace monitoring, thereby increasing operational security and improving the efficiency of such energy-intensive systems.

\section*{Acknowledgment}
\label{section:acknowledgment}
\noindent
This work was supported by the TITAN-FOURNOS (Tube Skin Temperature Monitoring System for Process Furnaces) project BI-2019/00024, which was financed by the Basque Industry 4.0 program of the Basque Government (SPRI) and is a collaborative project between Petronor, Petronor Innovation, University of the Basque Country, and Vicomtech Foundation.

\section*{Copyright Information}
\noindent
This manuscript version is made available under the CC BY-NC-ND 4.0 license. URL: 
\url{https://creativecommons.org/licenses/by-nc-nd/4.0/deed.en} \\
\includegraphics[width=7em]{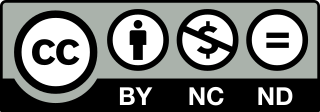}

\appendix
\section{Sensitivity analysis}\label{app:sensitivity}
See Figs. \ref{fig:sensitivityM2}, \ref{fig:sensitivityM3} and \ref{fig:sensitivityM4}.

\newcommand{\fw}{0.40\linewidth}
\begin{figure*}[!htb]
    \centering
    \subfloat[Wavelength $\lambda$ influence on measured temperature $T_s$ ]{\includegraphics[width=\fw]{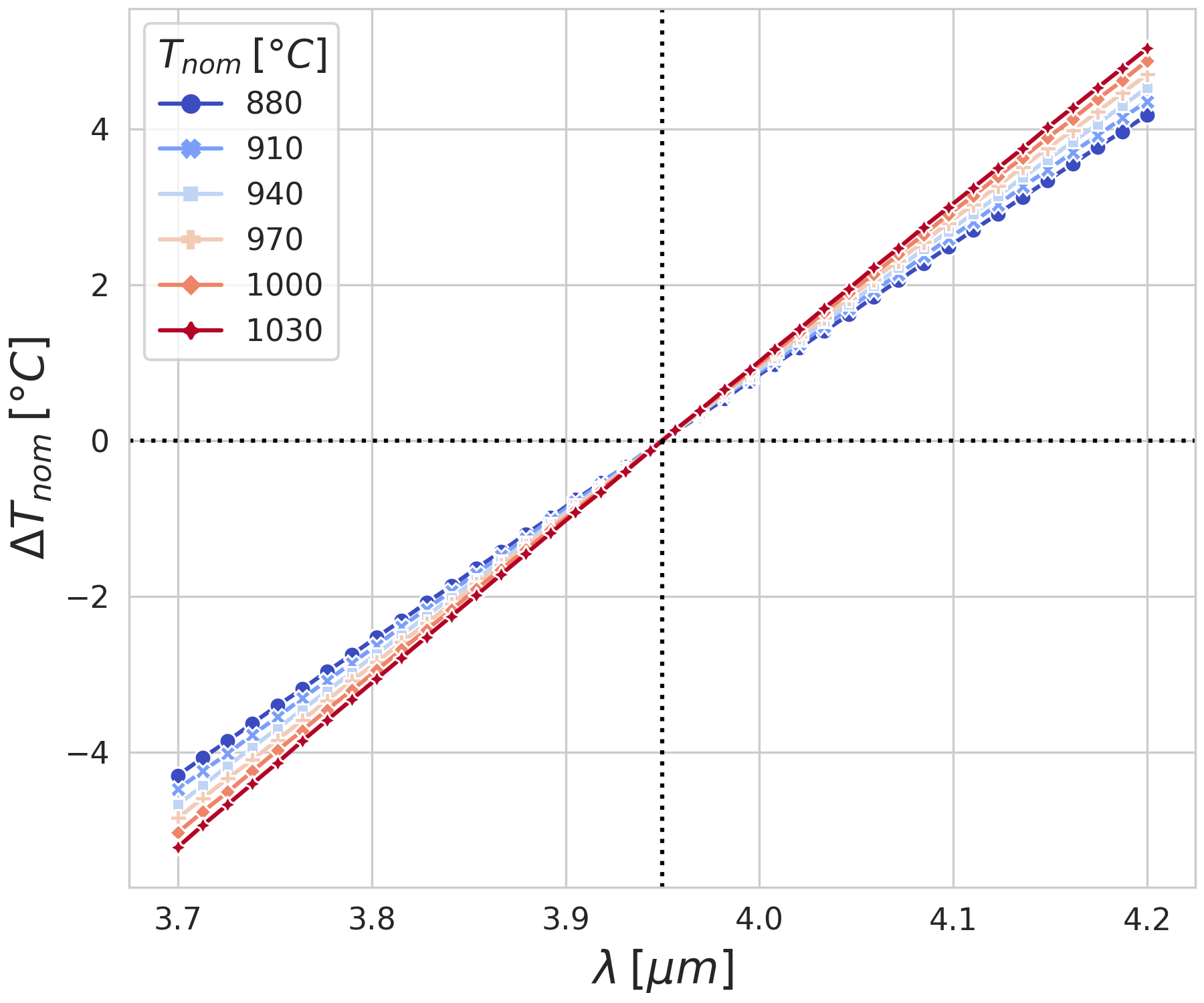} \label{fig:sensitivityM2A} }
    \hspace{1.5cm}
    \subfloat[Emissivity $\varepsilon$ influence on measured temperature  $T_s$]{\includegraphics[width=\fw]{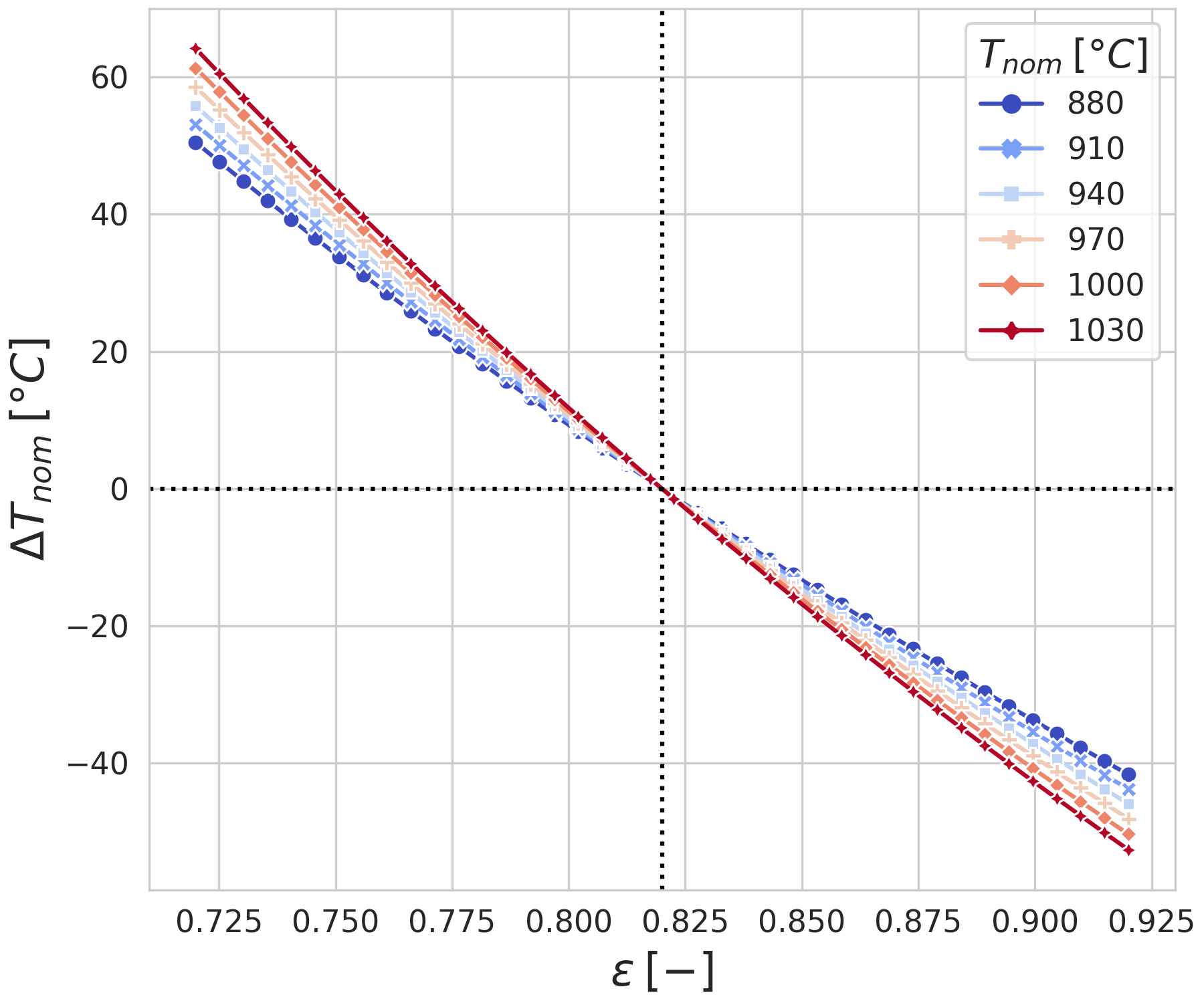} \label{fig:sensitivityM2B} }
    \caption{Model B sensitivity analysis: influence of model parameters ($\lambda$, $\varepsilon$) on the tube measured  temperature. Nominal values: $\lambda=3.95\mu m$, $\varepsilon=0.82$.
    Each line represents a different nominal temperature $T_s=[880, 910, 940, 970, 1000, 1030]\, \degree C$} 
    \label{fig:sensitivityM2}
\end{figure*}

\begin{figure*}[!htb]
    \centering
    \subfloat[Wavelength $\lambda$ influence on measured temperature $T_s$]{\includegraphics[width=\fw]{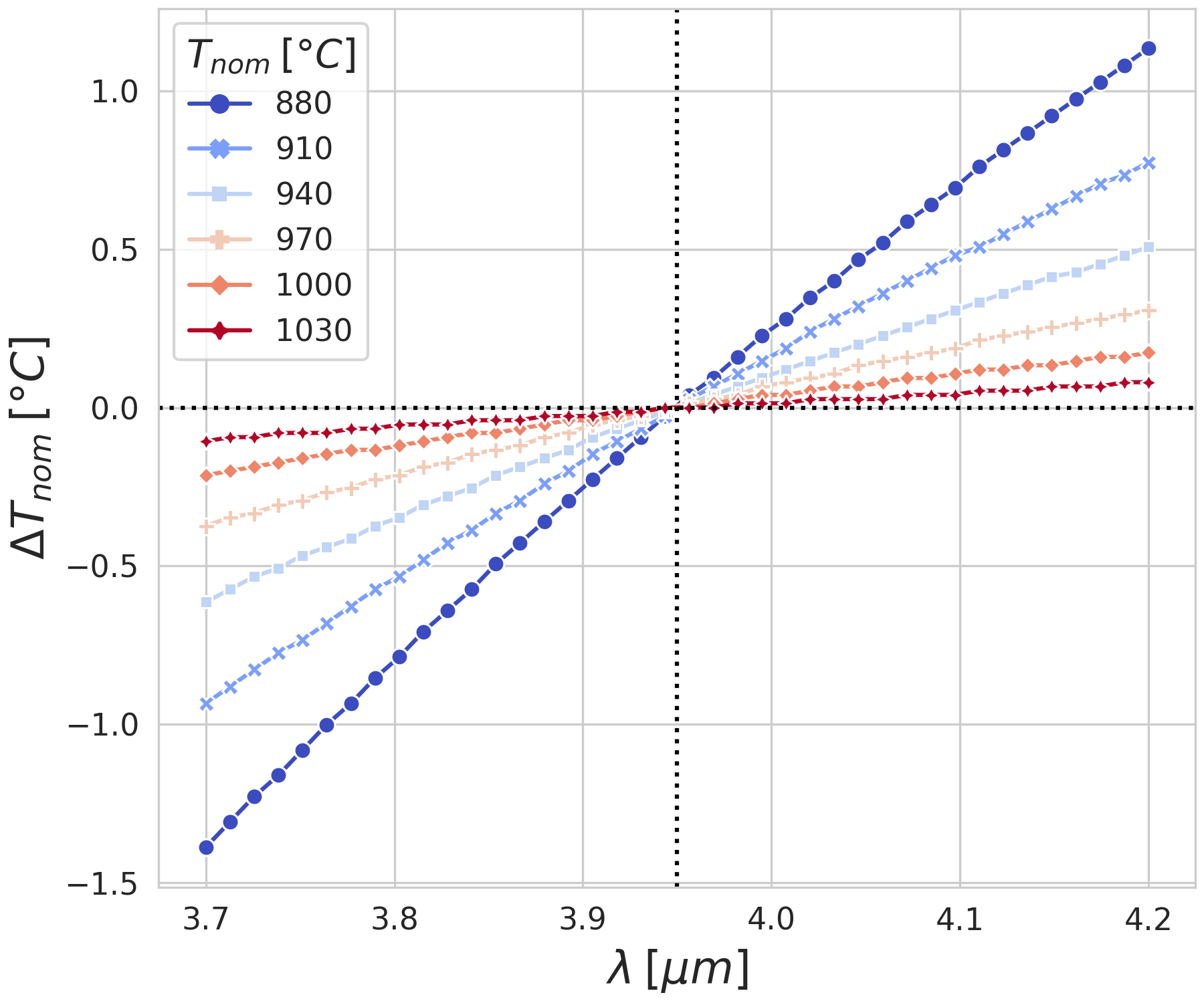} \label{fig:sensitivityM3A} }
    \hspace{1.5cm}
    \subfloat[Emissivity $\varepsilon$ influence on measured temperature $T_s$]{\includegraphics[width=\fw]{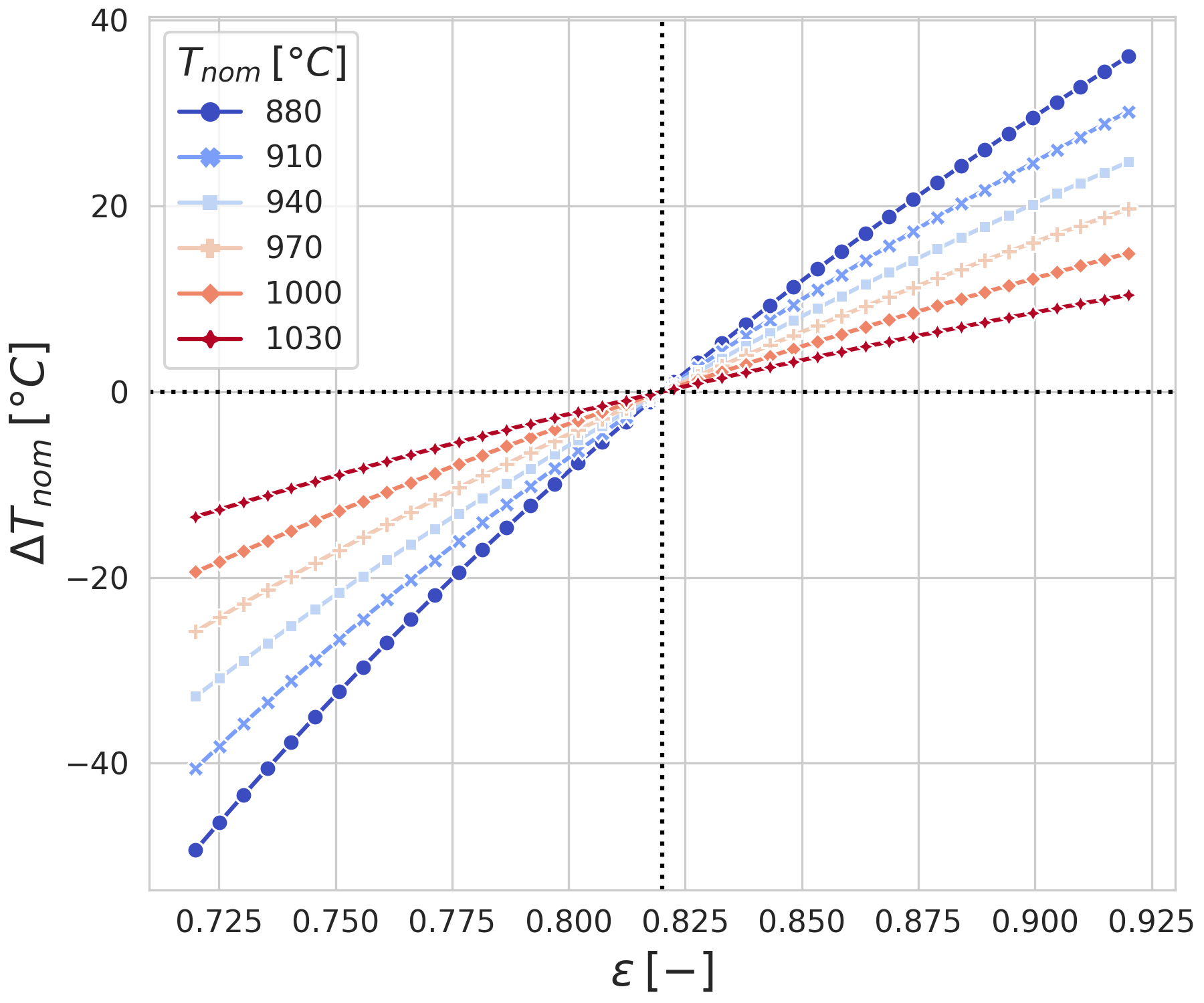} \label{fig:sensitivityM3B} } \\
    \hspace{0.75cm}
    \subfloat[Wall temperature $T_w$ influence on measured temperature $T_s$]{\includegraphics[width=\fw]{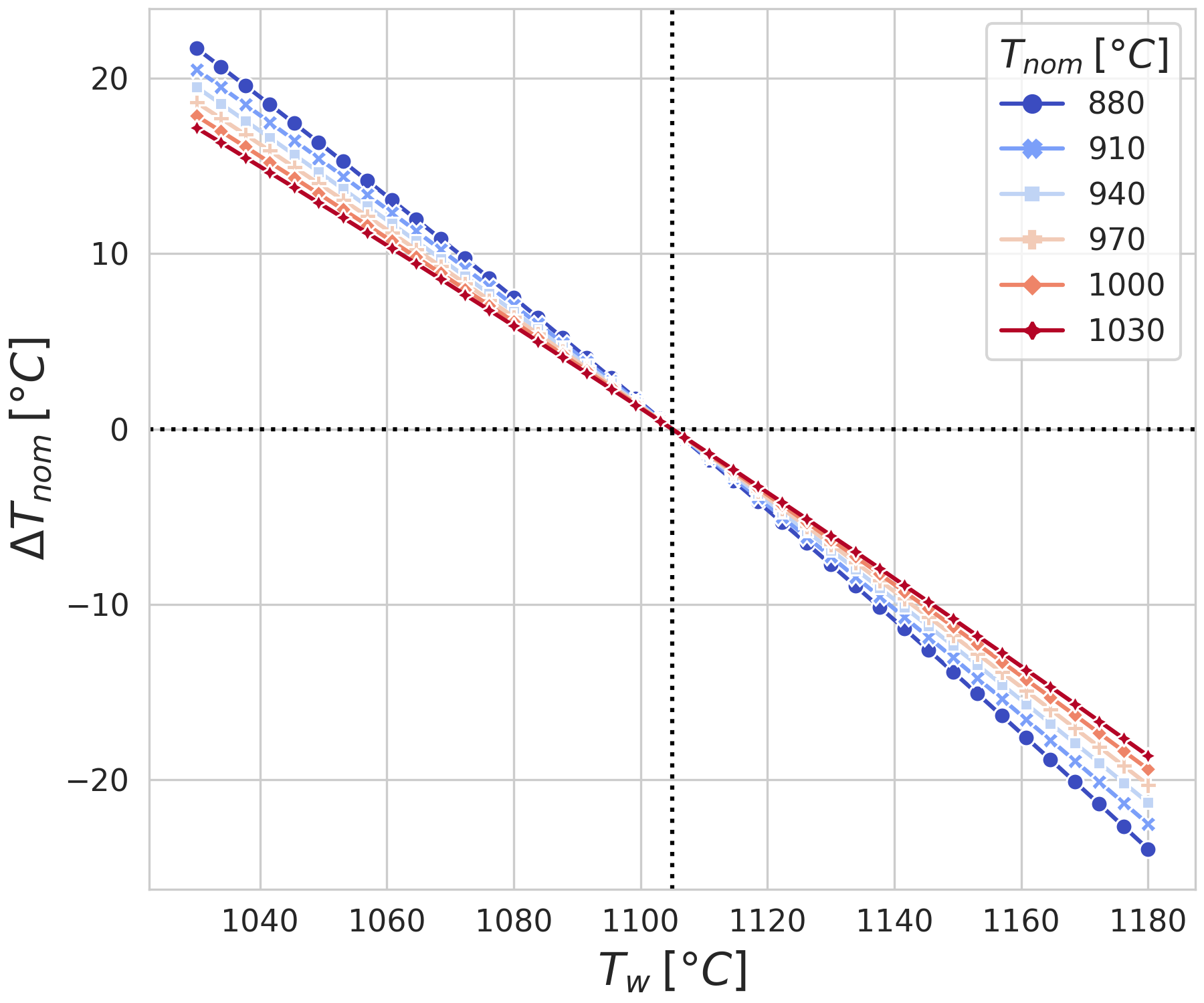} \label{fig:sensitivityM3C} }
    \hfill
    \caption{Model C sensitivity analysis: influence of model parameters ($\lambda$, $\varepsilon$, $T_w$) on the tube measured  temperature. Nominal values: $\lambda=3.95\mu m$, $\varepsilon=0.82$, $T_w=1105 \degree C$.
    Each line represents a different nominal temperature $T_s=[880, 910, 940, 970, 1000, 1030]\, \degree C$} 
    \label{fig:sensitivityM3}
\end{figure*}

\begin{figure*}[!htb]
    \centering
    \subfloat[Wavelength $\lambda$ influence on measured temperature $T_s$]{\includegraphics[width=\fw]{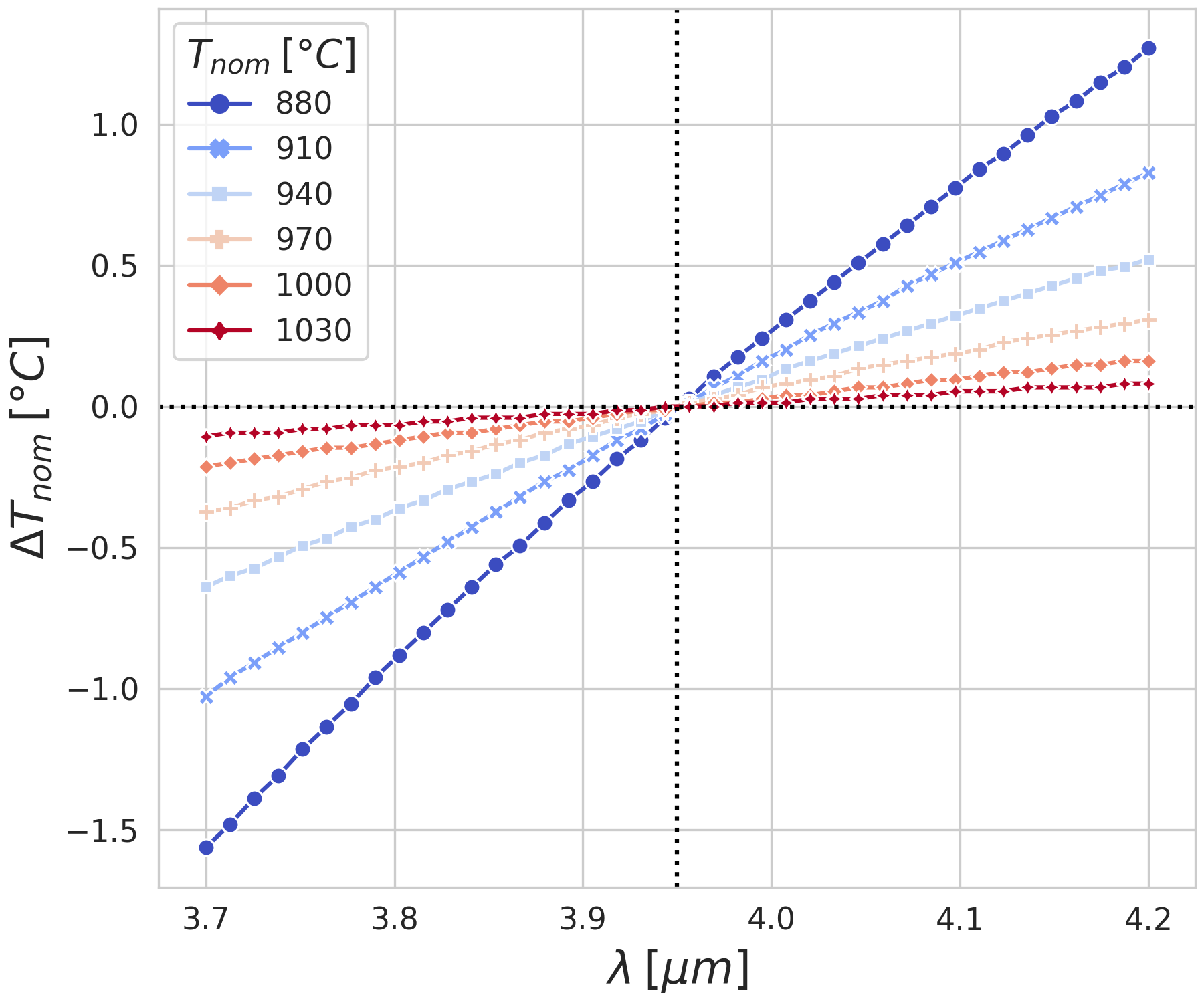} \label{fig:sensitivityM4A} }
    \hspace{1.5cm}
    \subfloat[Emissivity $\varepsilon$ influence on measured temperature $T_s$]{\includegraphics[width=\fw]{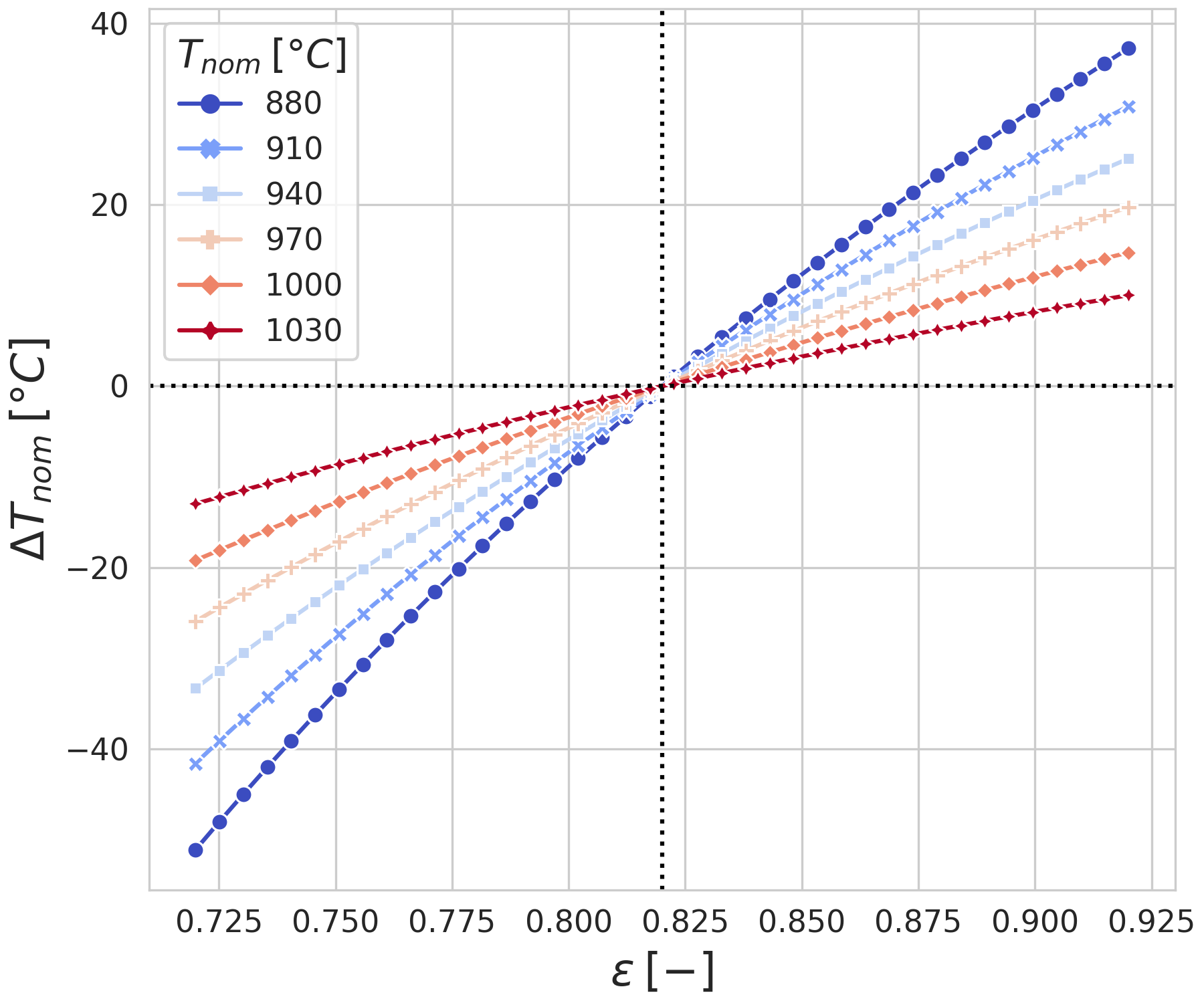} \label{fig:sensitivityM4B} } \\
    \subfloat[Wall temperature $T_w$ influence on measured temperature $T_s$]{\includegraphics[width=\fw]{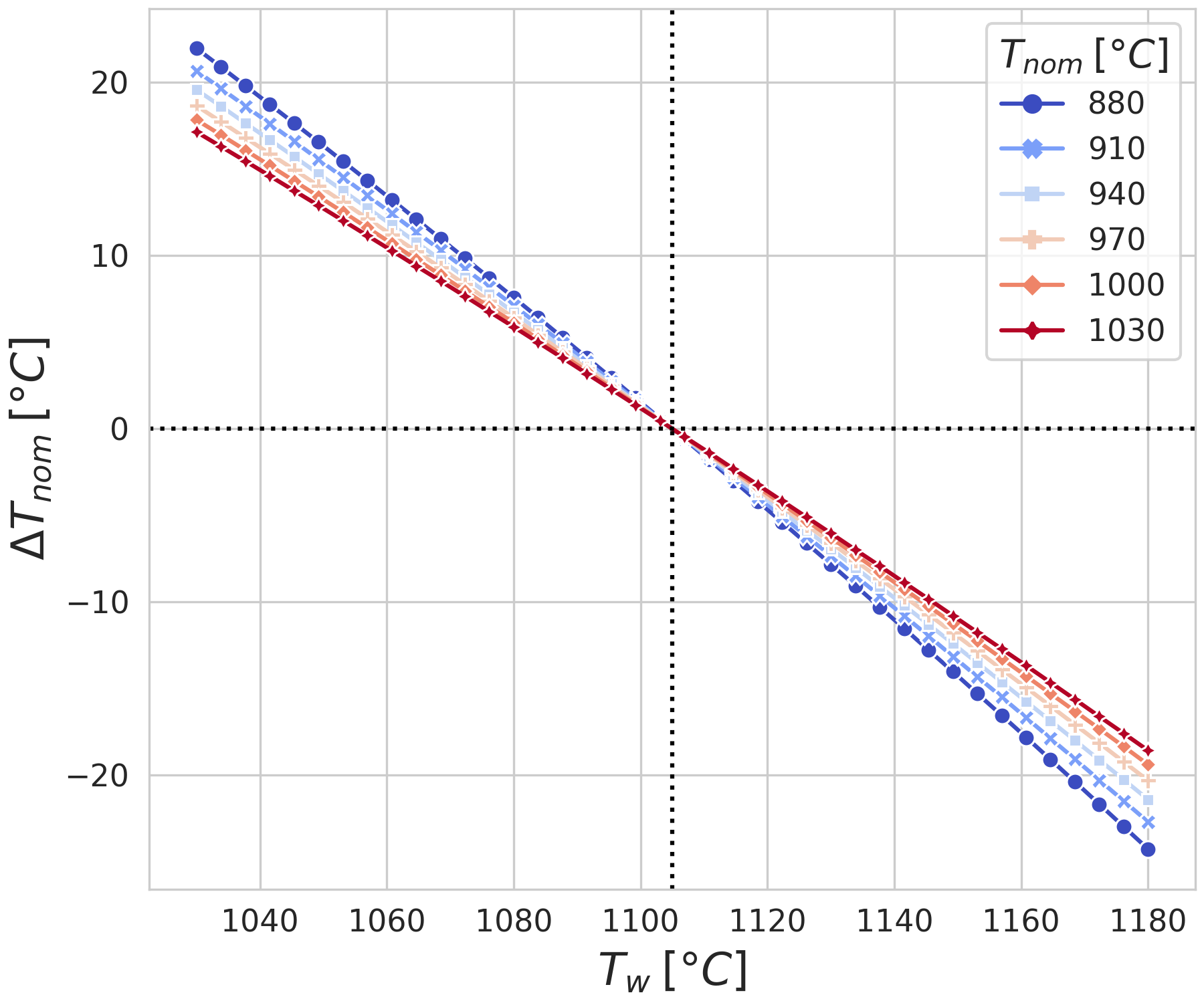} \label{fig:sensitivityM4C} } 
    \hspace{1.5cm}
    \subfloat[Fuel gas absorption $\alpha$ influence on measured temperature $T_s$]{\includegraphics[width=\fw]{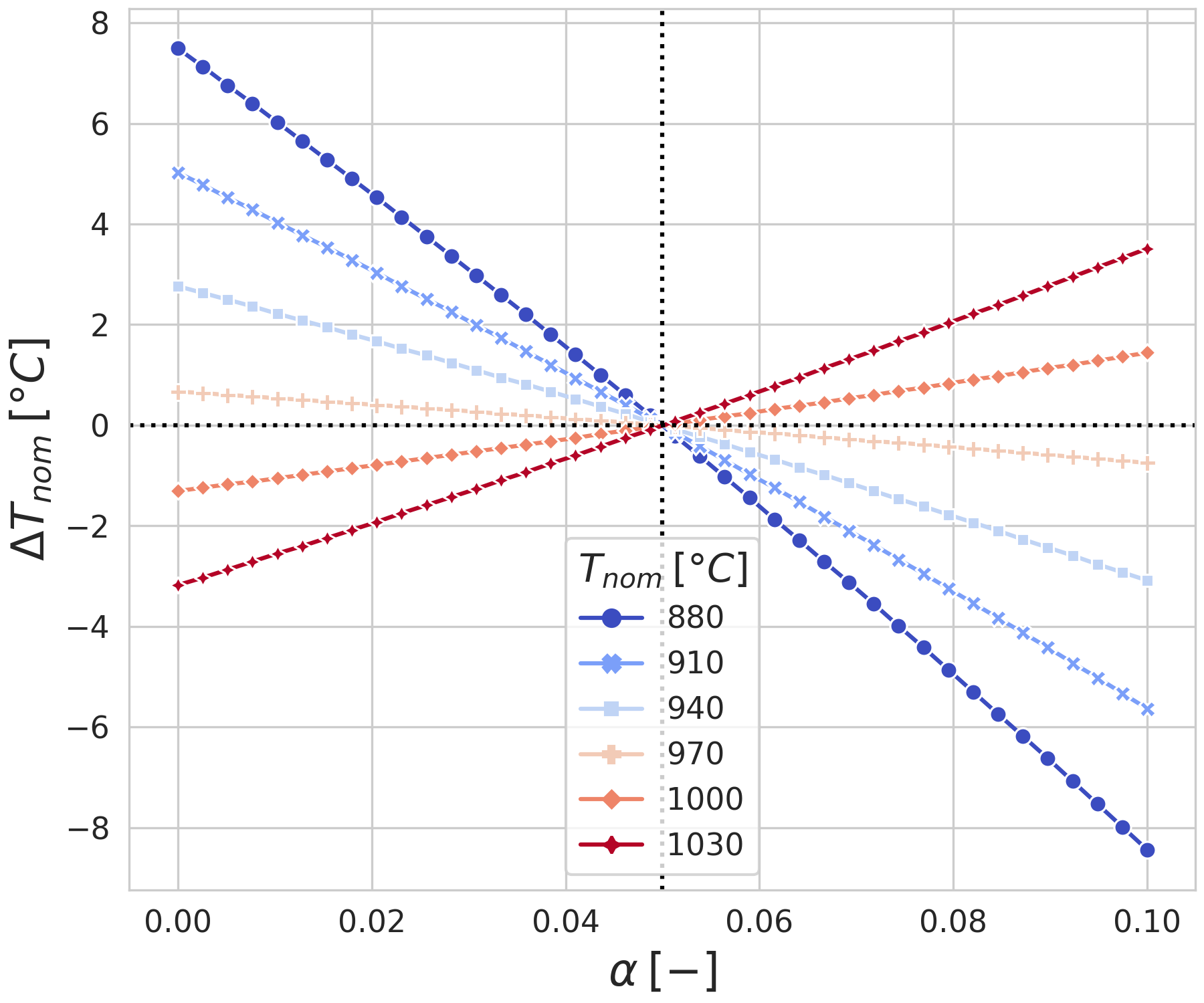} \label{fig:sensitivityM4D} } \\
    \hspace{0.75cm}
    \subfloat[Fuel gas temp. $T_w$ influence on measured temperature $T_s$]{\includegraphics[width=\fw]{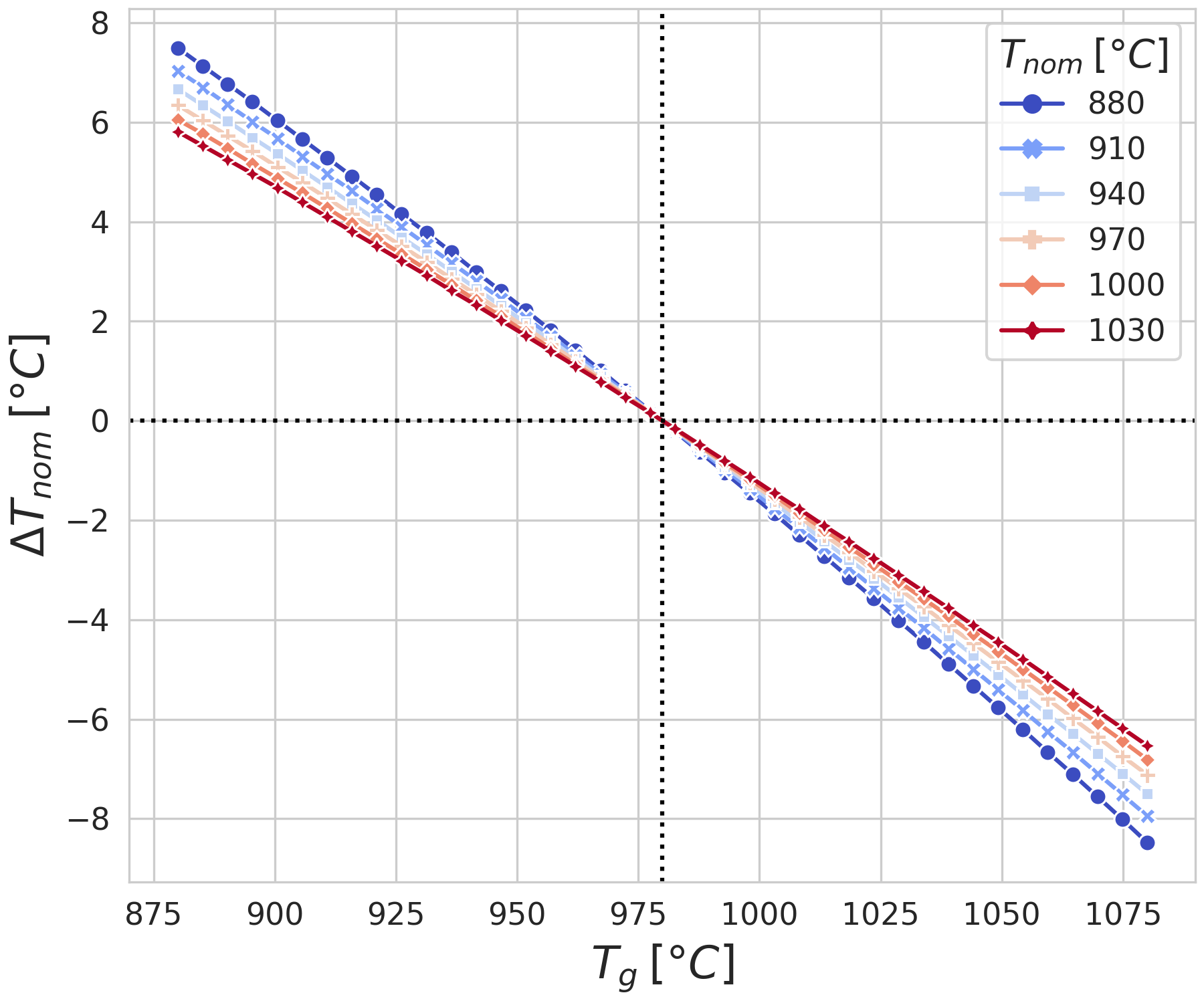} \label{fig:sensitivityM4E} }
    \hfill
    \caption{Model D sensitivity analysis: influence of model parameters ($\lambda$, $\varepsilon$, $T_w$, $\alpha$, $T_g$) on the tube measured  temperature. Nominal values: $\lambda=3.95\mu m$, $\varepsilon=0.82$, $T_w=1105 \degree C$, $\alpha=0.1$, $T_g=980 \degree C$.
    Each line represents a different nominal temperature $T_s=[880, 910, 940, 970, 1000, 1030]\, \degree C$} 
    \label{fig:sensitivityM4}
    \vspace{1cm}
\end{figure*}

\begin{figure*}[!htb]
    \centering
    \subfloat[Model B uncertainty analysis: influence of parameters ($\lambda$, $\varepsilon$) ]{\includegraphics[width=\fw]{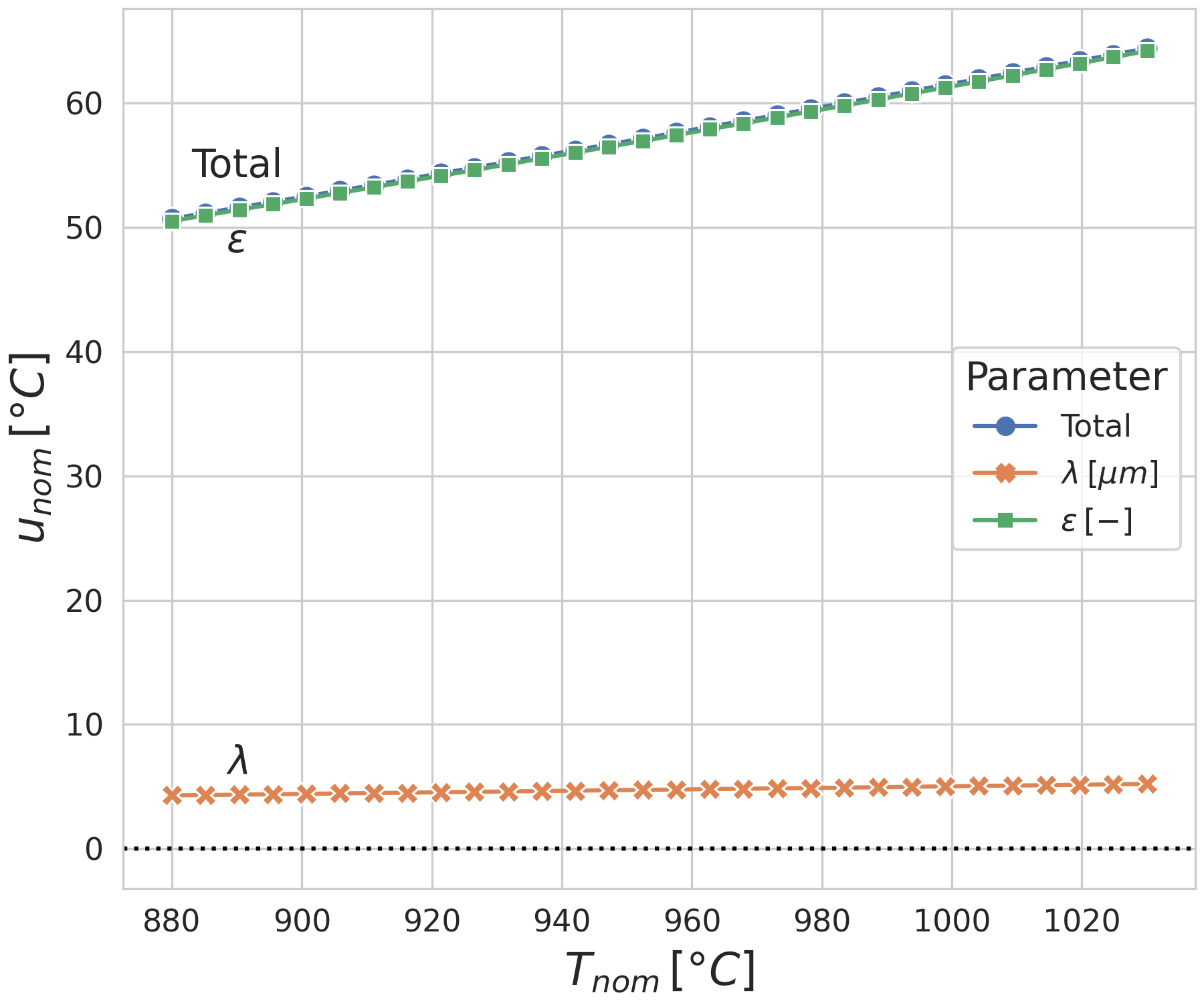} \label{fig:uncertaintyM2} }
    \hspace{1.5cm}
    \subfloat[Model C uncertainty analysis: influence of parameters ($\lambda$, $\varepsilon$, $T_w$) ]{\includegraphics[width=\fw]{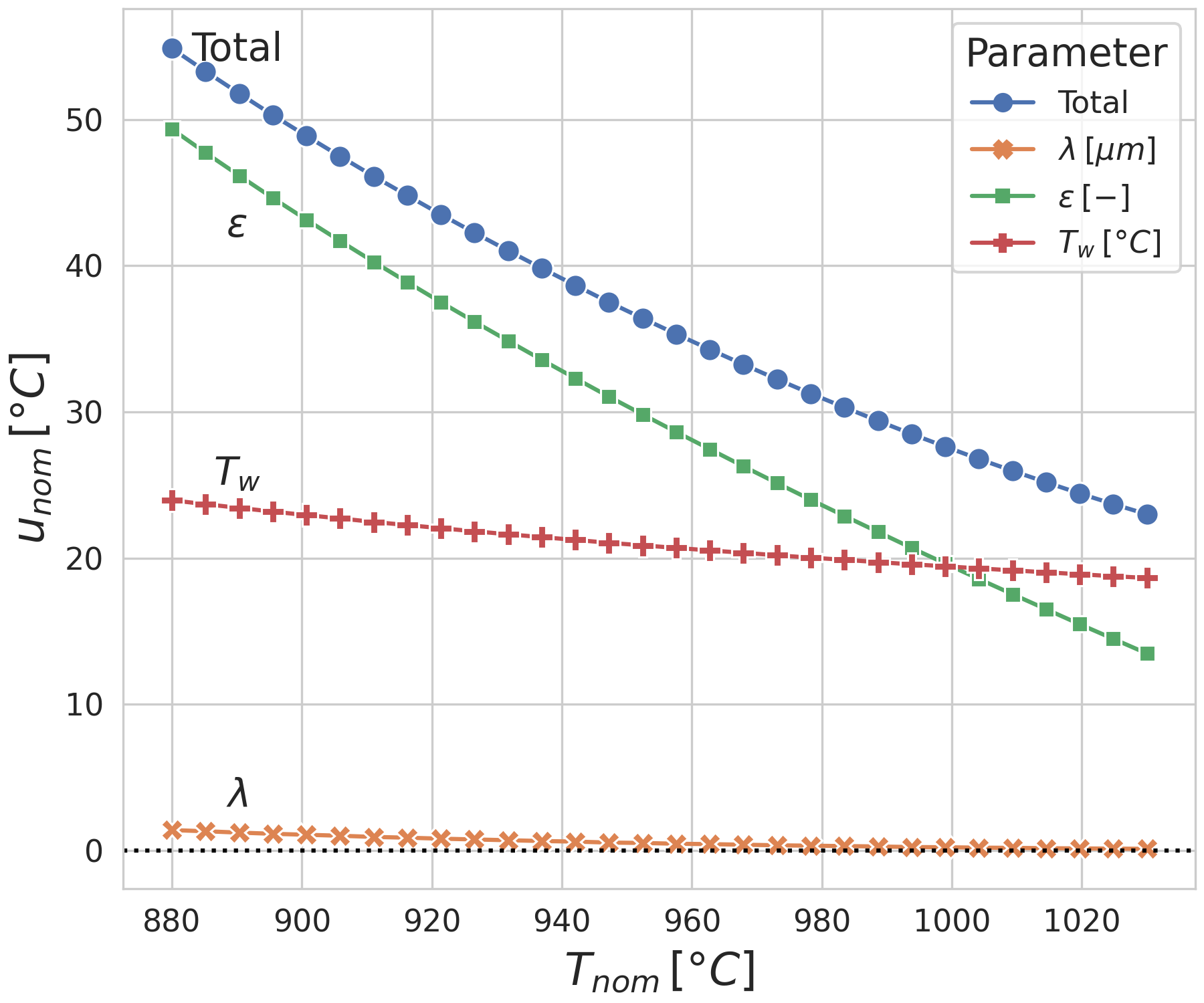} \label{fig:uncertaintyM3} } \\
    \hspace{0.75cm}
    \subfloat[Model D uncertainty analysis: influence of parameters ($\lambda$, $\varepsilon$, $T_w$, $\alpha$, $T_g$) ]{\includegraphics[width=\fw]{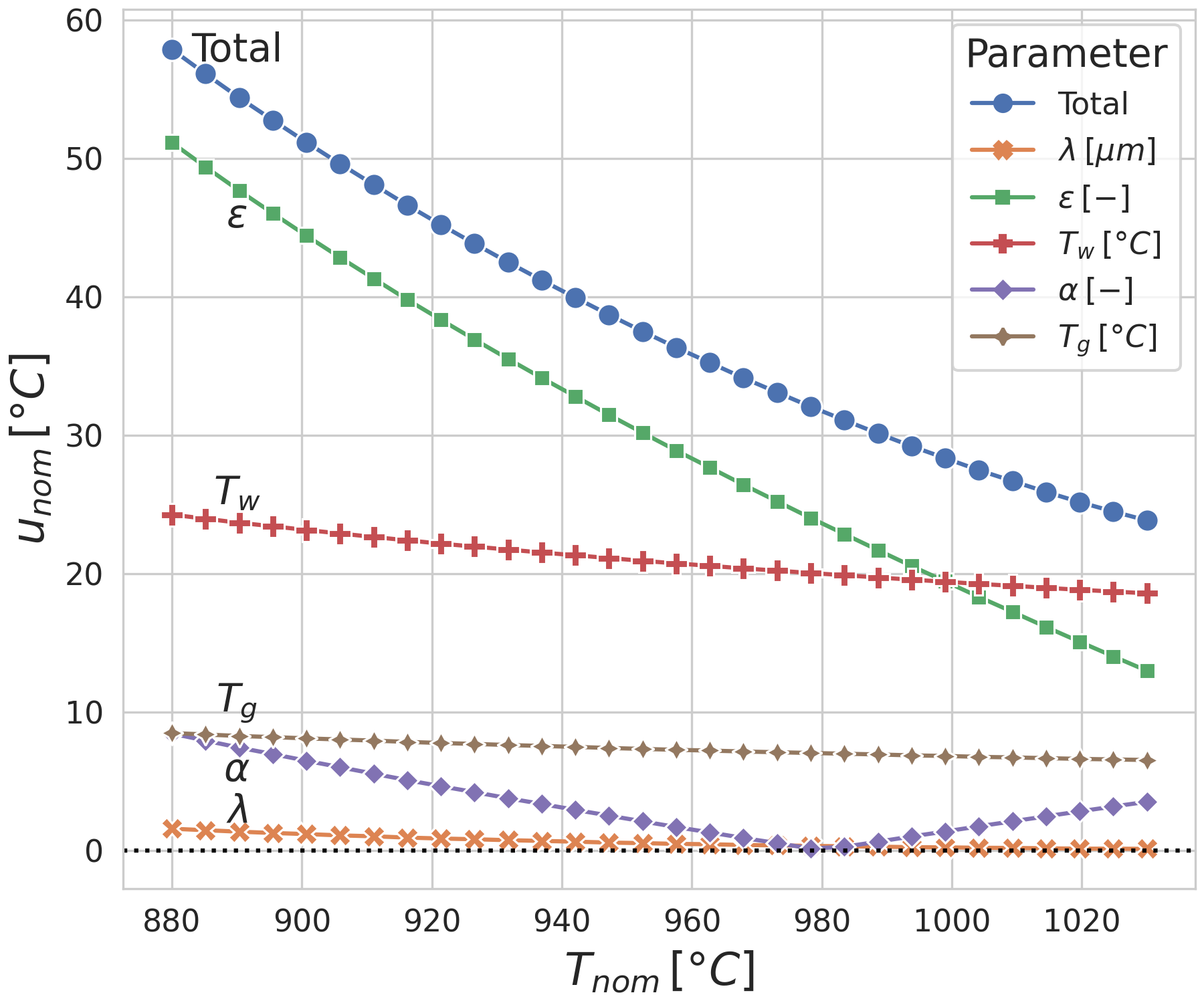} \label{fig:uncertaintyM4} } 
    \hfill
    \caption{Uncertainty analysis: influence of model parameters ($\lambda$, $\varepsilon$, $T_w$, $\alpha$, $T_g$) on the tube measured temperature uncertainty $u_{nom}$. Uncertainties add in quadrature to compute the combined uncertainty: $u_{c} = \sqrt{\sum_{i} u_{i}^2}$. Expanded uncertainty $U=k \cdot u_{c}$ is excluded due to its disproportionate scale ($k=1.96$).}
    \label{fig:uncertainty}
\end{figure*}

\bibliography{references}

\begin{thebibliography}{10}
\expandafter\ifx\csname url\endcsname\relax
  \def\url#1{\texttt{#1}}\fi
\expandafter\ifx\csname urlprefix\endcsname\relax\def\urlprefix{URL }\fi
\expandafter\ifx\csname href\endcsname\relax
  \def\href#1#2{#2} \def\path#1{#1}\fi

\bibitem{osti_1248754}
A.~F. Sabine~Brueske, Caroline~Kramer,
  \href{https://www.osti.gov/biblio/1248754}{Bandwidth study on energy use and
  potential energy savings opportunities in u.s. petroleum refining} (6 2015).
\newline\urlprefix\url{https://www.osti.gov/biblio/1248754}

\bibitem{life_intro}
C.~Moss, P.~Barrien, A.~Walczynski, Life management of refinery furnace tubing,
  International Journal of Pressure Vessels and Piping 77~(2) (2000) 105--112.

\bibitem{ir_intro}
P.~Pregowski, G.~Goleniewski, W.~Komosa, W.~Korytkowski, S.~Zwolenik, {Heating
  medium absorption and emission as factors in thermographic investigations of
  petrochemical furnaces}, in: D.~D. Burleigh, R.~B. Dinwiddie (Eds.),
  Thermosense XXXI, Vol. 7299, International Society for Optics and Photonics,
  SPIE, 2009, pp. 131 -- 141.

\bibitem{ir_intro2}
F.~Lopez, A.~Huot, {Advanced signal processing applied to thermographic
  inspection of petrochemical furnaces}, in: J.~de~Vries, B.~Oswald-Tranta
  (Eds.), Thermosense: Thermal Infrared Applications XLI, Vol. 11004,
  International Society for Optics and Photonics, SPIE, 2019, pp. 26 -- 32.

\bibitem{saunders}
P.~Saunders, Radiation thermometry: fundamentals and applications in the
  petrochemical industry, Vol.~78, SPIE press, 2007.

\bibitem{overheat}
{Babcock \& Wilcox}, Long-term overheat,
  \url{https://www.babcock.com/resources/learning-center/finding-the-root-cause-of-boiler-tube-failure}
  (2021).

\bibitem{thermometry}
D.~P. DeWitt, G.~D. Nutter, Theory and practice of radiation thermometry, John
  Wiley \& Sons, 1988.

\bibitem{thermocouples_2}
D.~D. Pollock, Thermocouples: theory and properties, CRC press, 1991.

\bibitem{thermocouple_1}
C.~R. Shaddix, Correcting thermocouple measurements for radiation loss: A
  critical review (7 1999).

\bibitem{Seebeck1}
D.~K. M.~Niffenegger, K.~Reichlin, The Change of the Seebeck Coefficient Due to
  Neutron Irradiation and Thermal Fatigue of Nuclear Reactor Pressure Vessel
  Steel and its Application to the Monitoring of Material Degradation, Paul
  Scherrer Institut, PSI, 2002.

\bibitem{Seebeck2}
A.~Patel, S.~K. Pandey,
  \href{http://dx.doi.org/10.1080/10739149.2016.1262396}{Automated
  instrumentation for high-temperature seebeck coefficient measurements},
  Instrumentation Science \& Technology 45~(4) (2016) 366–381.
\newblock \href {https://doi.org/10.1080/10739149.2016.1262396}
  {\path{doi:10.1080/10739149.2016.1262396}}.
\newline\urlprefix\url{http://dx.doi.org/10.1080/10739149.2016.1262396}

\bibitem{non-contact}
M.~Svantner, P.~Vacíková, M.~Honner, Non-contact charge temperature
  measurement on industrial continuous furnaces and steel charge emissivity
  analysis, Infrared Physics0 \& Technology 61 (2013) 20--26.

\bibitem{sota_ir}
R.~K. Weigle, {Applications of infrared thermography for petrochemical process
  heaters}, in: G.~R. Peacock, D.~D. Burleigh, J.~J. Miles (Eds.), Thermosense
  XXVII, Vol. 5782, International Society for Optics and Photonics, SPIE, 2005,
  pp. 100 -- 108.

\bibitem{Maldague1999}
X.~Maldague, Pipe inspection by infrared thermography, Materials Evaluation
  57~(9) (1999) 6 pp, cited By 16.

\bibitem{single-point}
O.~Struß, Transfer radiation thermometer covering the temperature range from
  -50\degree c to 1000\degree c (09 2003).

\bibitem{emissivity}
G.~Neuer, Spectral and total emissivity measurements of highly emitting
  materials, International Journal of Thermophysics 16~(1) (1995) 257--265.

\bibitem{gas_wavelength}
D.~Edwards, A.~Balakrishnan, Thermal radiation by combustion gases,
  International Journal of Heat and Mass Transfer 16~(1) (1973) 25--40.

\bibitem{dual-band}
K.~Irani, {Furnace wall-tube monitoring with a dual-band portable imaging
  radiometer}, in: D.~D. Burleigh, K.~E. Cramer, G.~R. Peacock (Eds.),
  Thermosense XXVI, Vol. 5405, International Society for Optics and Photonics,
  SPIE, 2004, pp. 221 -- 226.
\newblock \href {https://doi.org/10.1117/12.543644}
  {\path{doi:10.1117/12.543644}}.

\bibitem{ufpa}
N.~Oda, Uncooled bolometer-type terahertz focal plane array and camera for
  real-time imaging, Comptes Rendus Physique 11~(7-8) (2010) 496--509.

\bibitem{astme344}
ASTM International, West Conshohocken, PA, ASTM E344-20, Terminology Relating
  to Thermometry and Hydrometry, 2020th Edition (11 2020).

\bibitem{kirkup2006introduction}
L.~Kirkup, R.~B. Frenkel, An introduction to uncertainty in measurement: using
  the GUM (guide to the expression of uncertainty in measurement), Cambridge
  University Press, 2006.

\bibitem{saunders2008uncertainty}
P.~Saunders, J.~Fischer, M.~Sadli, M.~Battuello, C.~Park, Z.~Yuan, H.~Yoon,
  W.~Li, E.~Van Der~Ham, F.~Sakuma, et~al., Uncertainty budgets for calibration
  of radiation thermometers below the silver point, International Journal of
  Thermophysics 29~(3) (2008) 1066--1083.

\bibitem{corwin1994temperature}
R.~R. Corwin, A.~Rodenburgh, Temperature error in radiation thermometry caused
  by emissivity and reflectance measurement error, Applied optics 33~(10)
  (1994) 1950--1957.

\bibitem{liebmann2008infrared}
F.~E. Liebmann, M.~A.~C. Carrasco, Infrared uncertainty budget determination in
  an industrial application, in: Proc. of, 2008.

\bibitem{saltelli2002sensitivity}
A.~Saltelli, Sensitivity analysis for importance assessment, Risk analysis
  22~(3) (2002) 579--590.

\bibitem{willmott2016potential}
J.~R. Willmott, D.~Lowe, M.~Broughton, B.~S. White, G.~Machin, Potential for
  improved radiation thermometry measurement uncertainty through implementing a
  primary scale in an industrial laboratory, Measurement Science and Technology
  27~(9) (2016) 094002.

\bibitem{fpa_1}
G.~Machin, B.~Chu, High-quality blackbody sources for infrared thermometry and
  thermography between- 40 and 1000\degree c, The Imaging Science Journal
  48~(1) (2000) 15--22.

\bibitem{fpa_2}
N.~Oda, H.~Yoneyama, T.~Sasaki, M.~Sano, S.~Kurashina, I.~Hosako, N.~Sekine,
  T.~Sudoh, T.~Irie, Detection of terahertz radiation from quantum cascade
  laser using vanadium oxide microbolometer focal plane arrays, in: Infrared
  Technology and Applications XXXIV, Vol. 6940, International Society for
  Optics and Photonics, 2008, p. 69402Y.

\bibitem{fpa_3}
M.~Dem’yanenko, D.~Esaev, B.~Knyazev, G.~Kulipanov, N.~Vinokurov, Imaging
  with a 90 frames/ s microbolometer focal plane array and high-power terahertz
  free electron laser, Applied physics letters 92~(13) (2008) 131116.

\bibitem{land}
{AMETEK Land}, Fti-eb borescope,
  \url{https://www.ametek-land.com/obsolete-products/fti-eb-borescope} (2021).

\bibitem{relu}
A.~F. Agarap, \href{http://arxiv.org/abs/1803.08375}{Deep learning using
  rectified linear units (relu)}, CoRR abs/1803.08375 (2018).
\newblock \href {http://arxiv.org/abs/1803.08375} {\path{arXiv:1803.08375}}.
\newline\urlprefix\url{http://arxiv.org/abs/1803.08375}

\bibitem{adam}
D.~Kingma, J.~Ba, Adam: A method for stochastic optimization, International
  Conference on Learning Representations (12 2014).

\bibitem{de2020updated}
I.~G. de~Arrieta, T.~Ech{\'a}niz, R.~Fuente, J.~Campillo-Robles, J.~Igartua,
  G.~L{\'o}pez, Updated measurement method and uncertainty budget for direct
  emissivity measurements at the university of the basque country, Metrologia
  57~(4) (2020) 045002.

\end{thebibliography}

\end{document}